\documentclass[11pt,article,reqno]{amsart}
\usepackage{amsmath, amsthm, amsfonts}
\usepackage{amssymb,url}
\usepackage{mathtools}
\usepackage{hyperref}
\usepackage{amsaddr}
\usepackage[top=29mm,bottom=29mm,left=30mm,right=30mm]{geometry}
\usepackage{mathrsfs}
\usepackage[final]{graphicx}
\usepackage{ulem}
\usepackage{color}
\usepackage{mathtools}
\usepackage{graphicx}
\makeatletter
\def\section{\@startsection{section}{1}%
\z@{1\linespacing\@plus\linespacing}{1\linespacing}%
{\bf\centering}}
\def\subsection{\@startsection{subsection}{0}%
\z@{\linespacing\@plus\linespacing}{\linespacing}%
{\bf}}
\def\subsubsection{\@startsection{subsubsection}{0}%
\z@{\linespacing\@plus\linespacing}{\linespacing}%
{\bf}}
\makeatother

\makeatletter
\@addtoreset{equation}{section}
\makeatother

\newtheorem{theorem}{Theorem}[section]

\newtheorem{lemma}[theorem]{Lemma}
\newtheorem{proposition}[theorem]{Proposition}

\newtheorem{remark}[theorem]{Remark}
\theoremstyle{definition}
\newtheorem{definition}[theorem]{Definition}
\newtheorem{example}[theorem]{Example}
\newtheorem{assumption}[theorem]{Assumption}
\def\la{\lambda}

\def\bs{\mathbf{s}}
\def\bm{\mathbf{m}}
\def\bx{\mathbf{x}}
\def\bd{\mathbf{d}}
\def\bt{\mathbf{t}}
\def\bX{\mathbf{X}}

\def\scr{\mathscr}
\begin{document}
\title
{On the construction of  stationary processes and random fields }
\author{Jeonghwa Lee }

\address{Department of Mathematics and Statistics, University of North Carolina Wilmington, USA}
\email{leejb@uncw.edu}

\thanks{ 
 \\ \medskip
\noindent
2010 {\it MS Classification}: Primary {60G10}, {60G60}}

\begin{abstract}
We propose a new method to construct a stationary process and random field with a given decreasing covariance function and any one-dimensional marginal distribution. The result is a new class of stationary processes and random fields. The construction method utilizes a correlated binary sequence, and it allows a simple and practical way to model dependence structures in a stationary process and random field as its dependence structure is induced by the correlation structure of a few disjoint sets in the support set of the marginal distribution. Simulation results of the proposed models are provided, which show the empirical behavior of a sample path.   
\end{abstract}

\maketitle

\baselineskip 0.5 cm

\bigskip \medskip

%%%%%%%%%%%%%%%%%%%%%%%%%%%%%%%%%%%%%%%%%%%%%%
%% Please use \tableofcontents for articles %%
%% with 50 pages and more                   %%
%%%%%%%%%%%%%%%%%%%%%%%%%%%%%%%%%%%%%%%%%%%%%%
%\tableofcontents

%%%%%%%%%%%%%%%%%%%%%%%%%%%%%%%%%%%%%%%%%%%%%%
%%%% Main text entry area:

\section{Introduction}
 In time series, a stationary process is modeled by the moving average of independent innovations or the autoregression of past observations.
 These methods work well for various dependence structures, from an exponentially decaying covariance function to a power-law decaying covariance function, when the marginal distribution is a normal distribution. In practice, different types of marginal distribution, such as distributions with log-linear tails, appear in turbulence data and other empirical data from geology and mathematical finance  \cite{ Bar0, Bar1, Ebe}. Also, autocorrelation with more than one time scale was often observed in time series from financial data and turbulence data, and it is desirable to develop models for stationary processes where the one-dimensional marginal distribution can be chosen to fit the type of data at hand, while the dependence structure is flexible \cite{Bar2,Bar}.
In the past several decades, much progress has been made in constructing stationary processes with various marginal distributions.

  In \cite{Tav}, a Markovian process with exponential marginal distribution was developed. In \cite{Pit}, first-order stationary autoregressive models were introduced with non-Gaussian marginal distribution using a latent variable for the transition density. The method was extended for a continuous-time Markov model in \cite{Men}, and 
for broader marginal distributions in \cite{Con}. In \cite{Bar}, the sum of independent autoregressions was considered to construct a stationary process with a given marginal distribution and autocorrelation with two or more time scales.  A comprehensive account of other autoregressive models with non-Gaussian marginal distribution can be found in \cite{Bal}. 

Even though these models can account for a wide range of stationary processes, the marginal distributions are still restricted to self-decomposable distributions and the autocorrelations of the stationary processes are exponentially decaying since they are autoregressive/Markovian models with finite orders.

In a different direction,  
there has been an abundant development of modeling joint distributions with prescribed marginals (see \cite{Kot} for a comprehensive account). In \cite{Liu}, using earlier works of Morgenstern \cite{Mor} and Nataf \cite{Nat}, two methods were provided to model multivariate distributions with prescribed marginal distributions and covariances, and their potential uses in engineering were discussed. However, the applicability of the models is limited to low correlations or a restricted set of marginal distributions. 

In this paper, we propose a new method to construct a stationary process with any one-dimensional marginal distribution and a given covariance function that decays as fast as or slower than an exponential function. Our method utilizes a sequence of binary random variables whose covariance function is proportional to the given covariance function. The dependence structure of the stationary process will be induced by the binary process such that the states of the binary variable are linked to disjoint sets of the support set of the marginal distribution in the stationary process. This provides a different approach to model dependence in a stochastic process than existing moving average or autoregressive models as the dependence structure of the process is derived through a dependence structure among the disjoint sets in the marginal distribution of the process. When sets of values are known to be correlated and induce correlations in time series, our method can be a useful tool to model such data.
We also extend our method to construct a random field with a given marginal distribution and flexible covariance function.

In Section 2, we introduce some definitions and assumptions used in this paper. In Section 3 we propose a generalized Bernoulli process, a stationary binary sequence that can have a wide range of covariance functions.
In Section 4, we develop a method to construct a stationary process with a given one-dimensional marginal distribution and covariance function. We extend this method in Section 5 to propose a construction method for a random field with any one-dimensional marginal distribution and given covariance function.
In Section 6, simulations of some of our stationary processes and random fields are presented to graphically examine the behavior of a sample path and sample surface, followed by concluding remarks in Section 7.
All the proofs can be found in Section 8.

%in The work was extend in ... Some of them include non-Gaussian autoregressive models and the moving average of  resulting in markov stationary process non A different approach to construct stationary process Here we propose a new method to construct a stationary process with any marginal distribution and a convex covariance function.
%We also develop an estimation method for the parameters in such a probability model.

Throughout the paper, we will use bold font for a vector in two or more dimensions,  and
calligraphic font for a matrix.  We assume $i, i_0,i_1,\cdots \in \mathbb{Z},$ and $\bs, \bt, \bt_1,\bt_2,\cdots \in \mathbb{Z}^d,$ unless stated otherwise. For a random vector, its 
$i$th element is denoted with the subscript, e.g.,  $\bX(i)=(X_1(i), X_2(i),\cdots, X_d(i
))'.$
For identically distributed random variables, we suppress the index of a random variable for its expected value, i.e., if $\bX(i)=(X_1(i), X_2(i),\cdots, X_d(i)), i=1,2,\cdots,$ are i.i.d. random vectors, then we use $E(\bX),$ $ E(X_k),$ $ E(e^{ {\rm i} \bt' \bX}) $ for $E(\bX(i)), E(X_k(i))$ and $ E(e^{ {\rm i} \bt' \bX(i)}),$ respectively. 
For multiple integral with $\bx=(x_1,x_2,\cdots, x_d)$, we use notation $\int f(\bx) d\bx $ for $\int \cdots \int f(x_1,\cdots, x_d) dx_1 \cdots dx_d .$ For the union of disjoint sets, we use
$\sqcup$, for indicator function, we use $I_{\{ \cdot\}}$ or $I(\cdot),$ and  $[n]$ denotes the set $\{1,2,\cdots,n\}.$

\section{Preliminaries}
Here we present some operators and assumptions that will be used throughout the paper. Let $p\in(0,1),$ $C(\cdot)$ a positive function, and $C^*(\cdot)=C(\cdot)/p.$
The following operators were introduced in \cite{Lee}. 
\begin{definition} 
Define the following operation on a set $ B=\{i_0, i_1, \cdots,i_n\}\subset \mathbb{Z}$ with $i_0< i_1< \cdots<i_n. $
\[  L(B)= \prod_{j=1,\cdots,n}\Big(p+C^*(i_j-i_{j-1})\Big)  .\]
If $B=\emptyset$, define $ L(B):=1/p$, and if $|B|=1, L(B):=1. $ 
\end{definition}

 \begin{definition}
 Define for disjoint sets, $B,F\subset \mathbb{Z}$ with $ |F|=m>0,$  \begin{align*}
    D(B,F)=
    \sum_{j=0}^{|F|}\sum_{ \substack{F'\subset F\\ |F'|=j}}(-1)^{j} L(B\cup F').
 \end{align*} 
    If $F=\emptyset, D(B,F):=L(B).$
    
    \end{definition}
It will be shown in the next section that with the following assumptions, $C(\cdot)$ is a covariance function of a discrete-time, second-order stationary stochastic process.

 \begin{assumption} \label{assumption}
(Sufficient condition to be a covariance function)
\\{\it i)} $C: \mathbb{N} \to \mathbb{R}_+   $ is a decreasing function such that $C(x+1)/C(x) $ is non-decreasing as $x$ increases, and
\\ {\it ii)} $C(1)<p(1-p)$ and
    $C(2) > (p^2+C(1))^2/p-p^2$ for $ p\in (0,1).$
\end{assumption}

   Note that a decreasing function whose rate of decay is equal to or slower than exponential decay satisfies Assumption 2.3 ${\it i)}.$ 
In fact, many well-known convex, decreasing functions satisfy this condition including a power function, exponential function, etc.
\begin{remark}
{\it i)}
Any covariance function should be positive definite, and conversely, any positive definite function is a valid covariance function \cite{Cre}.
In the past several decades, there has been much progress in finding a class of covariance functions \cite{Por0, Por, Pos}.

For example, it is well known that there is a connection between positive definite functions and completely monotonic functions, from which Matérn  and Gneiting covariance functions are constructed. One can also find covariance functions through Fourier transform of positive and bounded measures by Bochner's theorem \cite{Boc}.\\
{\it ii)}
At first glance, our Assumption 2.3 {\it i)} may seem mild as any decreasing function that decays as slow as or slower than an exponential function satisfies this condition. However, according to Polya's criterion, any continuous, 
real-valued, even, decreasing function $f$ is positive definite if it is convex on $[0,\infty), f(0)=1,$ and $\lim_{t\to \infty} f(t)=0.$ 
\end{remark}

\section{A Generalized Bernoulli Process}
In \cite{Lee}, a generalized Bernoulli process (GBP) $\{\xi_i, i\in \mathbb{N}\} $ was introduced as a stationary binary sequence that can possess long-range dependence (LRD). The covariance function of GBP was a power function 
\[cov(\xi_i, \xi_{i+k}) = c|k|^{2H-2},\]
for some constants $c>0,H\in (0,1),$ 
and when $H\in (0.5, 1),$ GBP has LRD.
Here, we extend GBP to a more general stationary binary sequence with a covariance function that decays as fast as or slower than an 
exponential function.

Let $\xi_i, i\in \mathbb{Z},$ be a binary sequence and $C(\cdot)$ be a positive decreasing function satisfying Assumption 2.3. We will define GBP more broadly with the following probabilities. 
\begin{align*}
    &P(\xi_i =1)=p,  P(\xi_i =0)=1-p,\\
     &  P(\xi_{i_0}=1, \xi_{i_1}=1, \cdots, \xi_{i_n}=1)=
p\prod_{j=1}^{n}(p+C^*( i_j-i_{j-1} )), \end{align*} 
for $p\in (0,1),$ $i_0<i_1<\cdots<i_n,$ where $C^*(\cdot)=C(\cdot)/p.$
For any disjoint sets $B,F \subset \mathbb{Z},$  \begin{align*}
    &P( \cap_{i'\in F }\{\xi_{i'}=0\}  \cap_{i\in B}   \{\xi_{i}=1\} ) = \sum_{k=0}^{|F|}\sum_{\substack{F'\subset F\\ |F'|=k}} (-1)^{k}P(\cap_{i\in F'\cup B}\{\xi_{i}=1\} ), \text{ and }
\\ & P(\cap_{i'\in F }\{\xi_{i'}=0\})=1+\sum_{k=1}^{|F|}\sum_{\substack{F'\subset F\\ |F'|=k}} (-1)^{k}P(\cap_{i\in F'}\{\xi_{i}=1\} ).
\end{align*}
 
We can express the above probabilities more conveniently using the operator defined in the previous section. More specifically, if $D(B,F)\geq 0$ for any disjoint sets $B,F \subset \mathbb{Z}$,  GBP is well defined with  probabilities 
\begin{align}
     P(  \cap_{i\in B}\{\xi_{i}=1\} \cap_{i'\in F}\{\xi_{i'}=0\})&= pD(B,F)
\end{align}
(it is easy to check the consistency
conditions in the  Kolmogorov extension theorem are satisfied).
Especially, \[ P( \xi_i=\ell_1,\xi_j=\ell_2  )=\prod_{k=1,2}p^{\ell_k}(1-p)^{\ell_k} +(-1)^{\ell_1+\ell_2} C(|i-j|)\] for any $ \ell_1, \ell_2=0,1,$
therefore, it is a stationary process with covariance function $C,$
\[ Cov(\xi_i, \xi_j)=C(|i-j|), i\neq j.\]

The next theorem shows that under Assumption \ref{assumption}, GBP is well defined, i.e., $D(B,F) \geq 0$ for any disjoint sets $B,F\subset \mathbb{Z}.$

\begin{theorem} Under Assumption \ref{assumption},  GBP $\{\xi_i, i \in \mathbb{Z} \}$ is well-defined stationary binary sequence with $P(\xi_i=1)=p, P(\xi_i=0)=1-p,$ and  covariance function $C$.
\end{theorem}
We denote a GBP with a parameter $p$ and covariance function $C$ by  $\xi_i\sim GBP(p, C).$ Next, we show examples of covariance functions satisfying Assumption 2.3.

\begin{example}
A convex function, $C(x)=e^{-\theta x}, \theta>0,$ satisfies Assumption \ref{assumption} if $p\in(0,1), {\theta} \in ( -\ln ( (-p^2+p\sqrt{p-p^2+p^3})/(1-p) ), \infty).$ 
For $C(x)=ce^{-\theta x},$  if $\theta>0, c\in(0,p(1-p)),$  Assumption 2.3 is satisfied.
\end{example}

\begin{example}
A convex function, $C(x)=ce^{-\theta x^{\alpha}},  \alpha\in (0,1), c,\theta>0,$ satisfies Assumption 2.3 if the following two conditions are met.
\begin{align*}
   {\it i) }&\hspace{8pt} ce^{\theta}\in (p(1-p)/2, p(1-p)),\\
   {\it ii) }&\hspace{8pt} \alpha \in\Big(\log_2\Big({\frac{p(1-p)}{ce^{-\theta}}}\Big), 1\Big).
\end{align*}
\end{example}
\begin{example} A convex function, $C(x)=c_1\rho_1^{x}+c_2\rho_2^x, c_1,c_2>0, \rho_1,\rho_2\in (0,1)$, satisfies Assumption 2.3 if $c_1\rho_1+c_2\rho_2<p^{3/2}-p^2.$ 
    
\end{example}
\begin{example} In \cite{Lee}, a power function, $C(x)=c|x|^{2H-2}$, was used with the parameters 
 $p,H \in (0,1),$ and \[
    0\leq c<\min\{p(1-p), \frac{p}{2} ( -2p+2^{2H-2} +\sqrt{4p-p2^{2H}+2^{4H-4}})\},
\] which satisfies Assumption 2.3.
\end{example}

\section{Construction of a Stationary Process}
We will construct a  discrete-time stationary process with any one-dimensional marginal distribution and a  covariance function $C$ that satisfies Assumption 2.3. 
 More specifically, we define a multivariate stationary process $\{\bX(i) \in\mathbb{R}^d, i\in \mathbb{Z}\}$, $d \in \mathbb{N},$ where  $\bX(i)$ follows any given one-dimensional marginal distribution. For the sake of simplicity, we assume that the marginal probability distribution has pdf $f(\bx), \bx \in \mathbb{R}^d,$ but our method is easily extended with any probability space  $(\Omega, F, P)$ for a marginal distribution.  We utilize the GBP defined in Section 3 in the following construction method.

Let $\xi_i \sim GBP(p, C)$  and $\bX(i)= \bX^A(i) I_{\{\xi_i =1\}} +\bX^{A^c}(i) I_{\{\xi_i =0\}} $
where $\bX^A(i)\in\mathbb{R}^d, i\in \mathbb{Z},$ are i.i.d. random variables whose support set is $A\subset  \mathbb{R}^d$ such that $\int_A f({\bf x})d{\bf x}=p$ and the pdf of $\bX^A(i)$ is $f^A({\bf x}) =f({\bf x})/p, {\bf x}\in A  $. In the same way,  $\bX^{A^c}(i), i\in \mathbb{Z},$ are  i.i.d. random variables whose support set is $A^c$ and its pdf is 
$f^{A^c}( {\bf x})= f({\bf x})/(1-p), {\bf x}\in A^c$. Here, $f(\bx)$ can be any probability density function on $\bx=(x_1,x_2,\cdots,x_d) \in \mathbb{R}^d$.   Also, $\{\bX^A(i), i\in \mathbb{Z}\} $ and $\{\bX^{A^c}(i), i\in \mathbb{Z}\} $ are independent of each other and also independent of $\{\xi_i, i\in \mathbb{Z}\} $. 

\begin{theorem} \label{theorem 3.1}
  $\bX(i) \in \mathbb{R}^d, i\in \mathbb{Z},$ is a stationary process whose marginal probability density function is $f(\cdot)$ and variance-covariance  matrix is 
  \[ cov(\bX(i), \bX(j))=\scr{D} C(|i-j|) ,\] where
  \begin{align*}
    \scr{D}_{k,\ell}&=\Big(E(X_k^A)-E(X_k^{A^c})\Big)\Big(E(X_{\ell}^A)-E(X_{\ell}^{A^c})\Big)\\&=\frac{\int_A (x_k-\mu_k)f({\bf x}) d{\bf x} \int_A (x_{\ell}-\mu_{\ell})f({\bf x}) d{\bf x}} {p^2 (1-p)^2} \end{align*} for $k,\ell=1,2,\cdots,d,$  and ${\bf \mu}=(\mu_1,\mu_2,\cdots,\mu_d)'=E(\bX),$ i.e.,  $\mu_k=\int\cdots \int_{\mathbb{R}^d} x_k f({\bf x}) d{ \bf x}. $
  
\end{theorem}

For $q\in \mathbb{N},$
let  $ \bX^q(i)=( (X_1(i))^q,\cdots,(X_d(i))^q  )' $,  $ (\bX^{A})^q(i)= (  (X_1^A(i))^q,\cdots, (X_d^A(i))^q  )',$ and $ (\bX^{A^c})^q(i)= (  (X_1^{A^c}(i))^q,\cdots, (X_d^{A^c}(i))^q  )',$ then we have the following result which can be proved in a similar way to Theorem \ref{theorem 3.1}.
\begin{proposition}
For $q\in\mathbb{N},$  
  \[ cov(\bX^q(i), \bX^q(j))=\scr{D}^{(q)} C(|i-j|) ,\] where $\scr{D}^{(q)}={\bf d}^{(q)} {\bf d}^{(q)'} , {\bf d}^{(q)}= E( (\bX^{A})^q)-E((\bX^{A^c})^q),  $
i.e.,

\begin{align*} \scr{D}^{(k)}_{k,\ell}&= \Bigg( \frac{\int_A x_k^qf({\bf x}) d{\bf x}}{p} - \frac{\int_{A^c} x_{k}^qf({\bf x}) d{\bf x}}{1-p} \Bigg)\Bigg( \frac{\int_A x_{\ell}^qf({\bf x}) d{\bf x}}{p} - \frac{\int_{A^c} x_{\ell}^qf({\bf x}) d{\bf x}}{1-p} \Bigg)\\&=
\frac{\int_A(x_k^q-\mu_k^{(q)})f({\bf x}) d{\bf x} \int_A(x_{\ell}^q-\mu_{\ell}^{(q)})f({\bf x}) d{\bf x} }{p^2(1-p)^2},
\end{align*}   and  $\mu_k^{(q)}=\int x_k^q f({\bf x}) d{\bf x} $ for $k,\ell=1,2,\cdots,d.$
\end{proposition}

 \begin{proposition}
 For any set $B_1, B_2\subset \mathbb{R}^d,$
  \[ cov(I_{\{\bX(i)\in B_1\}}, I_{ \{\bX(j)\in B_2 \} })=d^* C(|i-j|) ,\] where
  \[d^*=\prod_{k=1,2}\Bigg( \frac{\int_{A\cap B_k} f({\bf x}) d{\bf x}}{p} - \frac{\int_{A^c \cap B_k } f({\bf x}) d{\bf x}}{1-p} \Bigg).\]  
\end{proposition}
\begin{remark}
Proposition 4.3 indicates that the covariance between $ I_{\{\bX(i)\in B_1\}}$ and $ I_{ \{\bX(j)\in B_2 \} }  $, for any $i, j \in \mathbb{Z},$ is maximized with $cov( I_{\{\bX(i)\in B_1\}},  I_{ \{\bX(j)\in B_2 \} } ) =C(|i-j|)$ when $B_1=B_2=A \text{ or } B_1=B_2=A^c.$
On the other hand, the covariance is minimized with  $cov( I_{\{\bX(i)\in B_1\}},  I_{ \{\bX(j)\in B_2 \} } ) =-C(|i-j|)$ when $B_1=A, B_{2}=A^c, $ or $  B_2=A, B_{1}=A^c. $
\end{remark}
Note that the joint pdf of $\{\bX(i_j), j=1,2,\cdots,k \}$ is
\[f (\bX({i_1}),\cdots, \bX({i_k}) )=\sum_{B\sqcup F=[k]} \prod_{\ell\in B} I( \bX({i_{\ell}}) \in A) \prod_{\ell'\in F}I(  \bX({i_{\ell'}}) \in A^c)  
pD(B,F)\frac{\prod_{j=1}^kf(\bX({i_j}))}{p^{|B|}(1-p)^{|F|}}. \]
Especially, for $k=2,$
\[f(\bX(i), \bX(j)) =\sum_{\ell_1,\ell_2=0,1}I(\bX(i)\in A^{\ell_1}, \bX(j)\in A^{\ell_2})\bigg(1 +\frac{(-1)^{\ell_1+\ell_2}C(|i-j|)}{\prod_{k=1,2}p^{\ell_k}(1-p)^{1-\ell_k} }\bigg)
f(\bX(i))f(\bX(j)) \] where $A^1=A, A^0=A^c.$ The next proposition on the characteristic function of $\{ \bX(i), \bX(j) \}$ is easily followed.

\begin{proposition} For $\theta_j \in \mathbb{R}^d, j=1,2,$
    \[ E(e^{ {\rm i} (\theta_1' \bX(i)+ \theta_2' \bX(j)  )} )= E(e^{{\rm i} \theta_1' \bX} )E(e^{{\rm i} \theta_2' \bX} )+ C(|i-j|) \prod_{k=1,2} \Big(E\big( e^{{\rm i}  \theta_k' \bX^A  } \big)- E\big( e^{{\rm i}  \theta_k' \bX^{A^c}  } \big) \Big). \]
\end{proposition}
Now we are ready to show the characteristic function of $\{\bX(i_j),j=1,2,\cdots,k \} $ for any $ k\in\mathbb{N}.$
Let $K\subset [k], K\neq \emptyset.$
Define $K^{\circ}=\{j: j\in [k]/K, \min K<j<\max K \}$,  and $Q_K$ as the set of  partitions of $K$ such that for any $W\in Q_K$, it satisfies that $W=\{K_i, i=1,2,\cdots, |W|  \},$  where $ K_i$ are disjoint, $ |K_i|\geq 2, \sqcup_i K_i=K,$ and $ \max{K_i}<\min{K_j}$ for $i<j.$ Note that   $K\in Q_K.$
 Also, define $W^{\circ}=\sqcup K_i^{\circ}$ for $W=\{K_i,i=1,2,\cdots,|W|\}. $
Let $L^*(\{i_1,i_2,\cdots,i_k\})=p\prod_{j=2,3\cdots,k}C^*(i_j-i_{j-1} ) $ for any $i_1<i_2<\cdots<i_k, k\in\mathbb{N}/\{1\}, $
and for $W=\{K_i, i=1,2,\cdots |W| \}\in Q_K,$ $L^*(W)$ is defined as $ \prod_{i=1}^{|W|} L^*( \{ i_j: j\in K_i\})$.

\begin{theorem} For $\theta_j \in \mathbb{R}^d, i_j\in \mathbb{Z}, j=1,2,\cdots,k, $ and $k\in \mathbb{N}/\{1\},$
    \begin{align*}
 &E(e^{{\rm i} \sum_{j=1,2,\cdots,k} \theta_j' \bX(i_j)  } )=\prod_{j=1}^kE(e^{{\rm i} \theta_j'\bX })+\\&
 \sum_{\substack{K\subset [k]\\ |K|\geq 2 \\W\in Q_K}}L^*(W)\prod_{j\in W^{\circ}}E(e^{{\rm i} \bt_j'\bX^{A^c} }) \prod_{i\in K}(E(e^{{\rm i} \theta_i'\bX^A })  -  E(e^{{\rm i} \theta_i'\bX^{A^c} }) ) \prod_{\ell \in [k]/(W^{\circ}\cup K) } E(e^{{\rm i} \theta_{\ell}'\bX })  . \end{align*}
 
\end{theorem}
Theorem 4.6 implies that the multivariate stationary process $\{\bX(i), i\in \mathbb{Z}\}$ becomes a sequence of i.i.d. random vectors if and only if $C\equiv 0,$ since the characteristic function of $\bX^A$ cannot be the same as that of $ \bX^{A^c}$, as they have different support set.

 \section{Construction of a Stationary Random Field}
In this section, we define a stationary multivariate random field $\bX({\bf t})=(X_1(\bt), X_2(\bt),\cdots, X_d(\bt) )'\in\mathbb{R}^d $ on $ {\bf t}=(t_1,t_2)'\in \mathbb{Z}^2 $ with any one-dimensional marginal distribution and a covariance function satisfying Assumption 2.3.
 Let $\xi_{i}^1 , \xi_{i}^2, i\in \mathbb{Z},$ be two independent GBPs, $\xi^k\sim GBP(p_k, C_k), k=1,2.$

Let $f(\bx), \bx=(x_1,\cdots, x_d)\in \mathbb{R}^d,$ be a pdf of any one-dimensional marginal distribution, and 
define for ${\bf t}=(t_1,t_2)' \in \mathbb{Z}^2,$ 
\[ \bX({\bf t})=\sum_{\ell_1,\ell_2=0,1} \bX^{A^{\ell_1\ell_2}}({\bf t})  I({\xi_{t_1}^1=\ell_1 ,\xi_{t_2}^2=\ell_2 })\]
where $ A^{11}, A^{10}, A^{01}, A^{00} \subset \mathbb{R}^d$ are disjoint sets  such that $\int_{A^{\ell_1\ell_2}} f({\bf x}) d{\bf x} = \prod_{k=1,2} (p_k)^{ \ell_k} (1-p_k)^{ (1-\ell_k)}    $
  and $ \bX^{A^{\ell_1\ell_2} }({\bf t}) \in \mathbb{R}^d , {\bf t}\in \mathbb{Z}^2 , $ are i.i.d. with pdf 

\[  f^{A^{\ell_1\ell_2}}(\bx)=\frac{f(\bx) I( \bx\in A^{\ell_1\ell_2} ) }{\prod_{k=1,2} p_k^{ \ell_k} (1-p_k)^{ 1-\ell_k} } \] for $\ell_1,\ell_2=0,1.$
Also, the four sequences $ \{\bX^{A^{\ell_1\ell_2}}({\bf t})  , {\bf t} \in \mathbb{Z}^2\}, \ell_1,\ell_2=0,1, $ are independent of each other, and also independent of $\{\xi_i^1,\xi_i^2, i\in\mathbb{Z}\}.$ 
\begin{theorem}
  $\{\bX(\bt) \in \mathbb{R}^d, {\bf t} \in \mathbb{Z}^2\}$ is a stationary random field whose marginal probability density function is $f(\cdot)$ and covariance function is
  \[C_X( {\bf t}-{\bf s} )=cov(\bX(\bt) , \bX(\bs))=\scr{M}_0C_1(|t_1-s_1|)C_2(|t_2-s_2|)+\sum_{j=1,2}\scr{M}_jC_j(|t_j-s_j|) \] for $\bt ,\bs \in \mathbb{Z}^2$ such that $t_i\neq s_i, i=1,2,$
  where $ \scr{M}_i=\bm_i\bm_i'$ for $ i=0,1,2,$ with \begin{align}
\bm_0&=E(\bX^{A^{10}})+E(\bX^{A^{01}})-E(\bX^{A^{11}})-E(\bX^{A^{00}}) ,\\
     \bm_1 &=
p_2\Big(E(\bX^{A^{11}})-E(\bX^{A^{01}})\Big) +(1-p_2)\Big(E(\bX^{A^{10}} )-E(\bX^{A^{00}}) \Big), \\
\bm_2&=p_1\Big(E(\bX^{A^{11}} )-E(\bX^{A^{10}})\Big) +(1-p_1)\Big(E(\bX^{A^{01}})-E(\bX^{A^{00}})\Big).
      \end{align}
      
      If $t_i\neq s_i$ and $ t_{i'}=s_{i'}$ for 
 $i,i'\in \{1,2\}$, 
      $C_X(\bt-\bs)=\scr{M}^*_{i}C_{i}(|t_{i}-s_{i}|)$
      with 
      \[\scr{M}^*_{1}=p_2{\bf u}^{1101} ({\bf u}^{1101})' +(1-p_2){\bf u}^{1000}({\bf u}^{1000})', \] and
      \[ \scr{M}^*_{2}= p_1 {\bf u}^{1110} ({\bf u}^{1110})' +(1-p_1){\bf u}^{0100}( {\bf u}^{0100}) ' \]
      where 
      \[{\bf u}^{\ell_1 \ell_2 \ell_3\ell_4}  =E(\bX^{A^{\ell_1 \ell_2}})-E(\bX^{A^{\ell_3\ell_4}})  \] for $\ell_j=0,1, j=1,2,3,4.$
If $\bt=\bs,$ $C_X(0)=var(\bX).$
\end{theorem}

Under certain conditions, 
the covariance function becomes simpler if we make (5.1) or (5.2-5.3) zero.
Let   $ \mu_i= \int_{\mathbb{R}^d} x_i f(\bx) d\bx ,\mu_i^{\ell_1\ell_2}=\int_{A^{\ell_1\ell_2}}x_if(\bx) d\bx 
  $ for $\ell_1, \ell_2=0,1, 
  i=1,2,\cdots, d,$ and $\mu_{i}^{(1)}= \mu_i^{11}+\mu_i^{10}$ and $ \mu_{i}^{(2)}= \mu_i^{11}+\mu_i^{01}.$  
Note that from (5.1-5.3),
if $\mu_{i'}^{(k)}=p_k \mu_{i'}$ for some ${i'}\in \{1,2,\cdots, d\}, k\in  \{1,2\}$, then $(\scr{M}_k)_{{i'},j}=(\scr{M}_k)_{j,{i'}}=0$ for all $j=1,2,\cdots, d.$ Moreover, if \[\mu_{i'}^{\ell_1\ell_2}= \mu_{i'}\prod_{k=1,2} p_k^{\ell_k} (1-p_k)^{1-\ell_k}, \] for all $\ell_1,\ell_2=0,1,$  
then $(\scr{M}_0)_{{i'},j}=(\scr{M}_1)_{{i'},j}=(\scr{M}_2)_{{i'},j}=(\scr{M}^*_{1})_{{i'},j}=(\scr{M}^*_{2})_{{i'},j}=0$ for all $j=1,2,\cdots,d.$ We summarize these results in the following propositions.
\begin{proposition}
    Let $\mu_i=\int_{\mathbb{R}^d} x_i f(\bx) d\bx=E(X_i(\bt))  ,$ and $\mu_{i}^{11}=\int_{A^{11}} x_if(\bx)d\bx$ for $i=1,2,\cdots, d.$ If  for some $i'$, \begin{align*}
\int_{A^{10}}x_{i'}f(\bx)d\bx&=p_1\mu_{i'}-\mu_{i'}^{11},\\ \int_{A^{01}}x_{i'}f(\bx)d\bx&=p_2\mu_{i'}-\mu_i^{11}, \text{ and }\\
\int_{A^{00}}x_{i'}f(\bx)d\bx&=(1-p_1-p_2)\mu_{i'} +\mu_{i'}^{11},\end{align*} 
    then $(\scr{M}_1)_{{i'},j}=(\scr{M}_2)_{{i'},j}=0$ for all $j=1,2,\cdots,d,$   and the covariance between  $X_{i'}(\bs)$ and $X_{j}(\bs+\bt)$ is
    \[ C_X( \bt )_{{i'},j} = (\scr{M}_0)_{{i'},j}C_1(|t_1|)C_2(|t_2|),\] for all $j=1,2,\cdots,d,$ and 
    $ \bt \in \mathbb{Z}^2$ such that $t_k \neq 0, k=1,2.$
    Moreover, if $\mu_{i'}^{11}=((p_1+p_2)/2-1/4)\mu_{i'}$, then $X_{i'}(\bs)$ is not correlated with any other variables $X_j(\bs+\bt)$, i.e., $(C_X(\bt))_{{i'},j}= {\bf 0},$ for all $\bt \in \mathbb{Z}^2$ such that $t_k \neq 0, k=1,2,$ and $ j=1,2,\cdots,d.$
\end{proposition}
\begin{proposition}
    If for some $i'\in\{1,2,\cdots,d\}, $\[\int_{A^{\ell_1\ell_2}} x_{i'}f(\bx) d\bx=\mu_{i'} \prod_{k=1,2} p_k^{\ell_k} (1-p_k)^{1-\ell_k}\] for all $\ell_1, \ell_2=0,1,$ then $(C_X(\bt))_{{i'},j}= {\bf 0}$ for all $\bt \in \mathbb{Z}^2/\{ {\bf 0}\} $ and $j=1,2,\cdots,d.$
\end{proposition}

We can extend the construction method for a random field on $\mathbb{Z}^2$ to a random field on $\mathbb{Z}^n.$ 
Let $\{\xi^1\},  \{\xi^2\}, \cdots, \{\xi^n \}$ be $n$ independent GBPs such that 
$\xi^k\sim GBP(p_k, C_k), k  \in [n].$ 

Define for ${\bf t}=(t_1,t_2,\cdots, t_n)' \in \mathbb{Z}^n,$ 
\begin{equation}
 \bX({\bf t})=\sum_{\ell_1,\ell_2, \cdots, \ell_n=0,1} \bX^{A^{\ell_1\ell_2\cdots \ell_n}}({\bf t})  I({\xi_{t_1}^1=\ell_1 ,\xi_{t_2}^2=\ell_2 },\cdots, \xi_{t_n}^n=\ell_n)\end{equation}
where $ A^{\ell_1\cdots \ell_n}, \ell_1, \ell_2,\cdots,\ell_n=0,1, $ are disjoint sets in  $\mathbb{R}^d$ such that $\int_{A^{\ell_1\ell_2\cdots\ell_n}} f({\bf x}) d{\bf x} = \prod_{k \in [n]} (p_k)^{ \ell_k} (1-p_k)^{ (1-\ell_k)}    $
  and $ \bX^{A^{\ell_1\ell_2\cdots \ell_n} }({\bf t}) \in \mathbb{R}^d , {\bf t}\in \mathbb{Z}^n, $ are i.i.d. with pdf 

\begin{equation}
 f^{A^{\ell_1\cdots \ell_n}}(\bx)=\frac{f(\bx) I( \bx\in A^{\ell_1\ell_2\cdots \ell_n} ) }{\prod_{k\in [n] } p_k^{ \ell_k} (1-p_k)^{ 1-\ell_k} } \end{equation} for $\ell_1,\ell_2, \cdots, \ell_n=0,1.$ 

\begin{theorem}
  $\{\bX(\bt) \in \mathbb{R}^d, {\bf t} \in \mathbb{Z}^n\}$ is a stationary random field with marginal probability density function  $f(\cdot)$ and covariance function 
  \begin{equation} {C}_X( {\bf t}-{\bf s} )=cov(\bX(\bt) , \bX(\bs))=\sum_{\substack{K \subset [n]/O_{(\bt-\bs)}\\ K\neq \emptyset }} \scr{M}_K^{(\bt-\bs)} \prod_{ k\in K} C_{k}(|t_{k}-s_{k}|) ,\end{equation}
  where \[ \scr{M}_K^{(\bt-\bs)}=\sum_{\substack{ \ell_{j'}=0,1
  \\ j'\in O_{(\bt-\bs)} }} \Big(\prod_{j' \in  O_{(\bt-\bs)} } p_{j'}^{\ell_{j'}}(1-p_{j'})^{1-\ell_{j'}} \Big)
  (\bm_K\bm_K'),\] and  \begin{align}
  \bm_K&= \sum_{\substack{\ell_{i'}=0,1\\ i'\in [n]/O_{(\bt-\bs)} }}(-1)^{|K|-\sum_{i\in K}\ell_i }  E\big(\bX^{A^{\ell_1\ell_2\cdots \ell_n}}\big)  \prod_{j\in [n]/(K\cup O_{(\bt-\bs)}) }p_j^{\ell_j}(1-p_j)^{1-\ell_j} .
      \end{align}
      Here, it is defined that $ \prod_{j\in \emptyset }p_j^{\ell_j}(1-p_j)^{1-\ell_j} =1$ and $  O_{(\bt-\bs)}= \{ j: t_j-s_j=0, j\in[n]\}.$ If $O_{(\bt-\bs)}=\emptyset, $ then $\scr{M}_K^{(\bt-\bs)}=\bm_K\bm_K'. $ 
  \end{theorem}

  Next, we will show when some terms in
  (5.6) become zero, which determines a set of combinations of indices $\{K: K\subset [n]\}$ that affect the covariance. 
For $\{\ell_i, i\in K\} \subset \{0,1\}^{|K|}, $ let \[A(\{\ell_i, i\in K\})  ={\bigcup}_{\substack{\ell_j=0,1\\ j\in [n] /K }}  A^{\ell_1\cdots \ell_n}\]
and $\mu_i=\int_{\mathbb{R}^n} x_i f(\bx) d\bx, i=1,2,\cdots, d.$
  \begin{proposition}
If for some $i'\in \{1,2,\cdots,d\}$ and non-empty set $ K'\subset \{1,2,\cdots,n \}$ we have
\begin{equation}
\int_{A(\{\ell_i, i\in K'\}) } x_{i'} f(\bx) d\bx = \mu_{i'}\prod_{i\in K'}p_i^{\ell_i}(1-p_i)^{1-\ell_i} \end{equation}      for all $\ell_i=0,1, i\in K',$ then
$(\scr{M}_{K}^{(\bt-\bs)})_{i',j}=0$ for all $j=1,2,\cdots,d, \bt,\bs\in \mathbb{Z}^n,$ and
 $K\subset K', K\neq \emptyset, $  such that $O_{(\bt-\bs)}\subset K'.$ 
  \end{proposition}
In the following remark, we consider how the condition (5.8) can be satisfied for a symmetric pdf $f.$
\begin{remark}
For a symmetric distribution $f,$ we can set $\{A^{\ell_1\cdots \ell_n}\}$ in the following way so that $A(\{\ell_i, i\in K'\})$  satisfies (5.8). Let $K'=\{i_1,i_2,\cdots,i_q\} \subset [n] , q\leq n .$ Let $f$ be  symmetric around ${\bf c}=(c_1,\cdots,c_d)', i.e., f({\bf x}+{\bf c})=f(-{\bf x}+{\bf c})$ for any $\bx=(x_1,\cdots,x_d)' \in \mathbb{R}^d$. For any $A_1^*\subset A^*=\mathbb{R}^{i'-1} \times (c_{i'},\infty) \times \mathbb{R}^{d-i'}$ such that $ \int_{A_1^*}f(\bx)d\bx=p_{i_1}/2 ,$ let $A_1=A_1^*\cup \{-\bx+{\bf c}: \bx+{\bf c}\in A_1^* \}.$ Also, let $A_0^*=A^*/A_1^*,$ and $A_0=A_0^*\cup \{-\bx+{\bf c}: \bx+{\bf c}\in A_0^* \}.$
Then, $\int_{A_1}f(\bx)d\bx=p_{i_1}, \int_{A_0}f(\bx)d\bx=1-p_{i_1},$
and \begin{align*}
\int_{A_1}x_{i'}f(\bx)d\bx&= \int_{ \bx+{\bf c}\in A_1^*}(x_{i'}+c_{i'})f(\bx+{\bf c})d\bx
+ \int_{ \bx+{\bf c}\in A_1^*}(-x_{i'}+c_{i'})f(-\bx+{\bf c})d\bx\\&
 = c_{i'}p_{i_1}.
\end{align*}
In a similar way, we can show that $\int_{A_0}x_{i'}f(\bx)d\bx=c_{i'}(1-p_{i_1}).$ Now, we choose any set $A_{11}^* \subset A_1^*$ such that $\int_{A_{11}^*}f(\bx) d\bx=p_{i_1}p_{i_2}/2,$ and $A_{01}^* \subset A_0^*$ such that $\int_{A_{01}^*}f(x) dx=(1-p_{i_1})p_{i_2}/2.$  Also, let $A_{10}^*=A_1^*/ A_{11}^* , A_{00}^*=A_0^*/ A_{01}^*, $ and $A_{\ell_1\ell_2}=A_{\ell_1\ell_2}^*\cup \{\bx+{\bf c}: -\bx+{\bf c}\in A_{\ell_1 \ell_2}^* \}, \ell_1,\ell_2=0,1. $ Then, it is easy to see that
$\int_{A_{\ell_1\ell_2}}f(\bx)d\bx=\prod_{j=1,2}p_{i_j}^{\ell_j}(1-p_{i_j})^{1-\ell_j}$ and
$\int_{A_{\ell_1\ell_2}} x_{i'} f(\bx)d\bx=c_{i'}\prod_{j=1,2}p_{i_j}^{\ell_j}(1-p_{i_j})^{1-\ell_j}.$ Using this method repeatedly, $A_{\ell_1\cdots \ell_q}$ can be defined such that
$A_{\ell_1\cdots \ell_q}^*\subset A_{\ell_1\cdots \ell_{q-1}}^*,  \int_{A_{\ell_1\cdots \ell_q}^*}f(\bx)d\bx=\prod_{j=1,\cdots,q}p_{i_j}^{\ell_j}(1-p_{i_j})^{1-\ell_j}/2,  A_{\ell_1\cdots \ell_q}=A_{\ell_1\cdots \ell_q}^*\cup \{\bx+{\bf c}: -\bx+{\bf c}\in A_{\ell_1\cdots \ell_q}^* \} $ for $\ell_1,\cdots ,\ell_q=0,1,$
which results in 
$\int_{A_{\ell_1\cdots \ell_q}}f(\bx)d\bx=\prod_{j=1,\cdots ,q}p_{i_j}^{\ell_j}(1-p_{i_j})^{1-\ell_j}$ and
$\int_{A_{\ell_1\cdots \ell_q}}x_{i'} f(\bx)d\bx=c_{i'}\prod_{j=1,\cdots, q}p_{i_j}^{\ell_j}(1-p_{i_j})^{1-\ell_j}.$
If $q=n,$ let $A^{\ell_1\cdots \ell_n}=A_{\ell_1^{\circ}\cdots \ell_n^{\circ}}$ where $\ell_j^{\circ}=\ell_{i_j} $ and (5.8) holds for $K'= \{1,\cdots, n\},$ and $(C_X(\cdot))_{i'v}\equiv 0$ for $v=1,\cdots,d.$  If $q<n,$ for $q<u \leq n,$ define ${i_u}=\min( [n]/\{i_1,\cdots, i_{u-1}\}),$ and find  $A_{\ell_1\cdots \ell_u}\subset A_{\ell_1\cdots \ell_{u-1}}$ such that $\int_{A_{\ell_1\cdots \ell_u}} f(\bx) d\bx =\prod_{j=1,\cdots,u}p_{i_j}^{\ell_j}(1-p_{i_j})^{1-\ell_j},$ but  $A_{\ell_1\cdots \ell_u}$ is not necessarily equal to $(A_{\ell_1\cdots \ell_u} \cap A^*)\cup \{ \bx+{\bf c}:  -\bx+{\bf c} \in (A_{\ell_1\cdots \ell_u} \cap A^*)\} $. Let $A^{\ell_1\cdots\ell_n}=A_{\ell_{1}^{\circ}\cdots\ell_{n}^{\circ}}$ where $\ell_j^{\circ}=\ell_{i_j},$ then (5.8) holds and
$(\scr{M}_{K}^{(\bt-\bs)})_{i',v}=0$ for all 
 $K\subseteq K', K\neq \emptyset, O_{(\bt-\bs)}\subset K'.$
Especially, if $K'=\{1,\cdots,n-1\},$ then for $\bt,\bs \in \mathbb{Z}^n $ such that $O_{(\bt-\bs)}= \emptyset,$
\[
 {C}_X( {\bf t}-{\bf s} )_{i'v}=  \prod_{ k\in [n] } C_{k}(|t_{k}-s_{k}|) \big(\bm_{[n]} \bm_{[n]}'\big)_{i'v}\] where
 \[\bm_{[n]}= \sum_{\substack{\ell_{i}=0,1\\ i\in [n]} }(-1)^{n-\sum_{i\in [n]}\ell_i }  E\big(\bX^{A^{\ell_1\ell_2\cdots \ell_n}}\big)  .\]
%In general, set $ A^{\ell_1\cdots\ell_n}$ disjoint sets such that $ \int_{A^{\ell_1\cdots\ell_n}}f(\bx)d\bx=\prod_{i=1,\cdots,n}p_i^{\ell_i}(1-p_i)^{1-\ell_i} $ for all $\ell_1,\cdots, \ell_n=0,1,$
%and $A^{\ell_1\cdots\ell_n}= (A^{\ell_1\cdots\ell_n}\cap A^*)\cup  \{ \bx+{\bf c}:  -\bx+{\bf c} \in (A_{\ell_1\cdots \ell_n} \cap A^*)\} $ for some $(\ell_1,\cdots, \ell_n) \subset \{0,1\}^n,$
%then (4.9) holds for 
\end{remark}

Next, we show how the condition (5.8) can be satisfied for a general pdf $f$ that is not necessarily symmetric.
\begin{lemma}
For any pdf $f$, $p\in (0,1),$ $A\subset  supp(f)=\{\bx: f(\bx)>0\} $, and a function $g $ such that $ \int_A |g(x)|f(x) dx <\infty,$  there exists $A_1\subset A $ such that $\int_{A_1} f(\bx) d\bx=p \int_{A} f(\bx) d\bx $ and $\int_{A_1} g(\bx) f(\bx) d\bx = p \int_A g(\bx) f(\bx) d\bx .$
\end{lemma}
Let $K'=\{i_1,i_2,\cdots,i_q\}\subset [n].$
By Lemma 5.7, for any pdf $f,$ 
we can find $A_1\subset supp(f)$ such that $\int_{A_1}f(\bx) d\bx=p_{i_1}$ and $ \int_{A_1}x_{i'} f(\bx) d\bx=p_{i_1}  \int x_{i'} f(\bx) d\bx. $ Let $A_0= A_1^c,$ then $\int_{A_0}f(\bx) d\bx=1-p_{i_1}$ and $ \int_{A_0}x_{i'} f(\bx) d\bx=(1-p_{i_1})  \int x_{i'} f(\bx) d\bx. $
Applying Lemma 5.7 repeatedly, we can find $A_{\ell_1 \cdots \ell_q} \subset supp(f)$ such that  $A_{\ell_1 \cdots \ell_q}\subset A_{\ell_1 \cdots \ell_{q-1}},$   \[\int_{A_{\ell_1 \cdots \ell_q}}f(\bx) d\bx=\prod_{j=1}^q p_{i_j}^{\ell_j} (1-p_{i_j} )^{1-\ell_j},\] and \[ \int_{A_{\ell_1 \cdots \ell_q}}x_{i'} f(\bx) d\bx= \prod_{j=1}^q p_{i_j}^{\ell_j} (1-p_{i_j} )^{1-\ell_j}  \int x_{i'} f(\bx) d\bx. \] Also, for $q<u\leq n,$ define  $ i_{u}=\min( \{1,\cdots, n\}/\{ i_1,\cdots, i_{u-1}\} )$  and we can find $A_{\ell_1 \cdots \ell_u} \subset A_{\ell_1 \cdots 
\ell_{u-1} }  $ such that 
$\int_{A_{\ell_1 \cdots \ell_u}}f(\bx) d\bx=\prod_{j=1}^u p_{i_j}^{\ell_j} (1-p_{i_j} )^{1-\ell_j}.$
Let $A^{\ell_1\cdots \ell_n}= A_{\ell_1^{\circ}\cdots \ell_n^{\circ} }$ where
$\ell_j^{\circ}=\ell_{i_j}$ for $j=1,\cdots, n.$ Then, by Proposition 5.5, this leads to 
\[  {C}_X( {\bf t}-{\bf s} )_{i'v}=\sum_{\substack{K \subset [n]/{O_{(\bt-\bs)}}\\ K \not \subset K'   \\K\neq \emptyset }} (\scr{M}_K^{(\bt-\bs)})_{i'v} \prod_{ k\in K} C_{k}(|t_{k}-s_{k}|)  \] for any $v=1,\cdots,d,$ and $\bt,\bs \in \mathbb{Z}^n$ such that $O_{(\bt-\bs)} \subset K'. $

Now we provide examples of our developed stochastic processes and random fields.

\begin{example}(Stationary process with exponential marginal distribution)\\
We use exponential distribution, $f(x)=\lambda e^{-\la x},  x>0, \la >0,$  a subset $A=(a,\infty) \subset \mathbb{R}_+ ,$   GBP$(p,C),$  $\xi_i, i\in \mathbb{Z},$ where $p=e^{-\la a} (=\int_A f(x) dx),$ and any decreasing function $C$ that satisfies Assumption 2.3. Let $X(i)=X^A(i)I(\xi_i=1) +  X^{A^c}(i)I(\xi_i=0 ),$  where 
 $X^A(i) \in \mathbb{R}, i\in \mathbb{Z} ,$ are i.i.d. random variables whose pdf is $f(x)I(x\in (a,\infty))/e^{-\la a} ,$ and   $X^{A^c}(i) \in \mathbb{R}, i\in \mathbb{Z} ,$ are i.i.d. random variables whose pdf is $f(x)I(x\in (0,a))/(1-e^{-\la a}).$
Then, $X(i), i\in \mathbb{Z},$
is a stationary process whose one-dimensional marginal distribution is exponential distribution, $X(i)\sim Exp(\la),$
and the covariance function is
\[cov(X(i), X(j))= \frac{a^2}{(1-e^{-\la a})^2} C(|i-j|),\]
for any $i\neq j.$
\end{example}

\begin{example}(Stationary process with uniform marginal distribution) \\
i)  With uniform distribution $f(x)= 1, x\in (0,1),$ and a subset  $A=(0,a), a\in (0,1),$ let  $X^A(i), i\in \mathbb{Z},$ be i.i.d. random variables whose pdf is  $ f^A(x)=1/a, x\in (0,a),$ and  $X^{A^c}(i), i\in \mathbb{Z},$ be i.i.d. random variables whose pdf is  $ f^{A^c}(x)=1/(1-a), x\in (a,1).$
Let $\{\xi_i\}$ be a GBP$(p,C)$ where $p=a$ and $C$ is any decreasing function that satisfies Assumption 2.3. 
Then, $X(i)(=X^A(i)I(\xi_i=1) +  X^{A^c}(i)I(\xi_i=0 ))$ is a stationary process with uniform marginal distribution and covariance function
\[ cov(X(i), X(j))= \frac{1}{4} C(|i-j|),\] for any $i\neq j,$
by Theorem 4.1 and $E(X^A)-E(X^{A^c})=a/2- (1+a)/2=-1/2 $. 
\\ ii) Using uniform distribution on $(0,1)^2,$ $f(\bx)=1, \bx \in (0,1)^2,$ a subset  $A=(0,a_1)\times (0,a_2), a_1,a_2\in (0,1),$ and GBP$(p,C)$  with $p=a_1a_2 $ and $C$ satisfying Assumption 2.3, 
define $\bX^A, \bX^{A^c},$ and $\bX(i)=\bX^A(i)I(\xi_i=1) +  \bX^{A^c}(i)I(\xi_i=0 )$ as in Section 4. It follows that $E(X_k^A)-E(X_k^{A^c}) = (a_k-1)/2(1-a_1a_2), k=1,2,$ and $\{\bX(i)\}$ is a bivariate stationary process with marginal uniform distribution on $(0,1)^2$ and covariance function
\[ cov(X_k(i), X_{\ell}(j))= \scr{D}_{k,\ell} C(|i-j|),\] where
\[\scr{D}_{k,\ell}= \frac{(a_k-1)(a_{\ell}-1)}{4(1-a_1a_2)^2 },\]  for any $i\neq j,$ and $ k,\ell=1,2.$
 \end{example}

\begin{example}(Binary random field in $ \mathbb{Z}^2 $)\\
   Let $\{\xi_i^1\}$ and $\{ \xi_i^2\}$ be GBPs with $(p_j,C_j), j=1,2.$ Let $X(\bt)=\xi_{t_1}^1 \xi_{t_2}^2$ for $\bt \in \mathbb{Z}^2,$ which is obtained if $X^{A^{11}}=1 , X^{A^{10}}=X^{A^{10}}=X^{A^{10}}= 0$ in (5.5). Then, $X(\bt)$ is 1 or 0 with $P(X(\bt)=1)=p_1p_2, P(X(\bt)=0)=1-p_1p_2,$ and
   \[cov(X(\bt), X(\bs))=\begin{dcases}
    \prod_{j=1,2}C_j(|t_j-s_j|)+p_1^2C_2(|t_2-s_2|)+p_2^2C_1(|t_1-s_1|) & \text{ if } t_j\neq s_j ,j=1,2, \\
    p_{j'}C_j(|t_j-s_j|)+p_j^2p_{j'}(1-p_{j'}) &\text{ if } t_{j'}=s_{j'}, t_j\neq s_j.
   \end{dcases} \]
   
 \end{example}

 \begin{example}(Random fields with Gaussian marginal distribution) \\
i) Here we use Gaussian pdf, $f(x)=\frac{1}{\sqrt{2\pi}}e^{-x^2/{2}}.$
 Let $A^{\ell_1 \ell_2}, \ell_1, \ell_2=0,1,$ be disjoint subsets of $ \mathbb{R}$ such that \begin{align}
    \mu^{\ell_1\ell_2}= (-1)^{\ell_1+\ell_2}a,\\
  \int_{A^{11}}f(x) dx:\int_{A^{10}}f(x) dx=\int_{A^{01}}f(x) dx: \int_{A^{00}}f(x) dx, \end{align} 
where  $\mu^{\ell_1\ell_2}=\int_{A^{\ell_1\ell_2} } xf(x) dx,  $ and for some $a\in (0, 1/(2\sqrt{2\pi})].$ 
For example, if $ A^{10}=(0,a'), A^{00}=(-a',0),$ for $a'>0$ such that $\int_{0}^{a'}x f(x) dx \leq 1/(2\sqrt{2\pi}).$ Then, one can find $b>a'$ such that for $A^{11}=(-\infty, -b)\cup(a',b), A^{01}=(b,\infty)\cup(-b, -a') ,$ we have $\mu^{11}  =\mu^{00},\mu^{01}  =\mu^{10} $ and $ \int_{A^{11}}f(x) dx=\int_{A^{01}}f(x)  dx,$ therefore, (5.9-5.10) are satisfied.  
Let $\{\xi^1\},\{\xi^2\} $ be GBPs with $(p_1,C_1), (p_2,C_2),$ respectively, where 
 $p_1= \int_{A^{10} \cup  A^{11} }f(x) dx, p_2= \int_{A^{01} \cup  A^{11} }f(x) dx,$ and $C_1,C_2$ are  decreasing functions satisfying Assumption 2.3. 
 Define $X^{A^{\ell_1\ell_2}}(\bt)$ and 
  $ X({\bf t}), \bt\in \mathbb{Z}^2 ,$ as (5.4-5.5).  It follows by Theorem 5.1 that $\{X(\bt) ,\bt \in \mathbb{Z}^2\}$ is a random field on $\mathbb{Z}^2$ with one-dimensional Gaussian marginal distribution and covariance function
 \begin{align}
     cov(X(\bt) , X(\bs)) =\begin{dcases}
         \scr{M}_0C_1(|t_1-s_1|)C_2(|t_2-s_2|) &  \text{ if } t_j\neq s_j, j=1,2, \\
         \scr{M}_j^*C_j(|t_j-s_j|)& \text{ if } t_j\neq s_j, t_{j'}=s_{j'},
     \end{dcases}\end{align} where $\scr{M}_0= a^2/(\prod_{i=1,2}p_i(1-p_i))^2,$
 and  $\scr{M}_j^*=2\scr{M}_0 p_{j'}(1-p_{j'}) .$ 
 \\ii) Use the same marginal distribution, $f(x)=\frac{1}{\sqrt{2\pi}}e^{-{x^2}/{2}},$ but with disjoint sets   $A^{\ell_1 \ell_2}\subset  \mathbb{R}, \ell_1, \ell_2=0,1,$ such that 
 \begin{align}
      \int_{A^{11}}f(x) dx:\int_{A^{10}}f(x) dx=\int_{A^{01}}f(x) dx: \int_{A^{00}}f(x) dx, \\
p_2 \mu^{10} +p_1 \mu^{01}= (1-p_1-p_2) \mu^{11},
 \end{align}
 where
$p_1=\int_{A^{11}\cup A^{10}} f(x) dx, p_2= \int_{A^{00}\cup A^{01}} f(x) dx$ 
 and 
 $ \mu^{\ell_1\ell_2}=\int_{A^{\ell_1\ell_2} } xf(x) dx.$ Let $\{\xi^1\},\{\xi^2\} $ be GBPs with $(p_1,C_1), (p_2,C_2),$ respectively, for any decreasing functions $C_1,C_2$ satisfying Assumption 2.3.  Define $X^{A^{\ell_1\ell_2}}({\bf t}) $ and $ X({\bf t})$ as (5.4-5.5). By Theorem 5.1, it follows that
 $ \{X({\bf t}), \bt\in \mathbb{Z}^2 \} $ is a stationary random field with one-dimensional Gaussian marginal distribution and covariance function
 \begin{align}
  cov(X(\bt) , X(\bs)) =\begin{dcases}
     \scr{M}_1C_1(|t_1-s_1|)+\scr{M}_2 C_2(|t_2-s_2|) & \text{ if } t_j\neq s_j, j=1,2, \\ \scr{M}_j^* C_j(|t_j-s_j|) &\text{ if } t_j\neq s_j, t_{j'}=s_{j'},  \end{dcases} \end{align} where $\scr{M}_j, \scr{M}_j^*  $ are computed as Theorem 5.1.
 For example, let $A^{10}=(-\infty, -a), A^{01}=(a, \infty)$ for any $a>0$ such that $z_a\leq .25,$ 
 $A^{11}=(-b, b)$ where $a>b>0$ such that $4(.5-z_b)(z_b-z_a)=z_a^2,$ and $A^{00}=\mathbb{R}/(A^{10}\cup A^{01} \cup A^{11}).$ Then, (5.12-5.13) is satisfied with $ \mu^{11}=\mu^{00}=0, \mu^{10}=-\mu^{01},$ therefore,
 $ \{X({\bf t}), \bt\in \mathbb{Z}^2 \} $ is a stationary random field with one-dimensional Gaussian marginal distribution and covariance function as (5.14) with $\scr{M}_0\equiv 0,$
   $\scr{M}_j=\Big(\int_a^{\infty}xf(x)dx/(p_j(1-p_j)) \Big)^2$  
 and $  \scr{M}_j^*=(\int_a^{\infty}xf(x)dx)^2( (1-p_j)^{-2}p_{j'}^{-1}+  (1-p_{j'})^{-1}p_{j}^{-2})$ for $ j=1,2.$
 \end{example}
 Note that in Example 5.11, we constructed stationary random fields $X$ that have one-dimensional  Gaussian marginal distribution, i.e.,  $X(\bt) \sim N(0,1)$ for any $\bt \in\mathbb{Z}^2$, however, they are not Gaussian random fields since $(X(\bt_1), X(\bt_2),\cdots, X(\bt_k)  ), k>1,$ does not follow multivariate normal distribution as we will show in the following.
 Let $\bt_j=(t_{j,1},t_{j,2})', $
 and 
 $T_i=\{ t_{j,i}: j=1,2,\cdots,k\}, $ for $i=1,2, j=1,2,\cdots,k.$
 Then the joint pdf of $\{\bX(\bt_j), j=1,2,\cdots,k \}$, for  $k\in \mathbb{N}, \{\bt_1,\cdots,\bt_k \}\subset \mathbb{Z}^2,$ is
\[f (\bx_{\bt_1},\bx_{\bt_2},\cdots, \bx_{\bt_k} )=\sum_{\substack{A_i^0\sqcup A_i^1=T_i \\ i=1,2}} \prod_{\substack{\bt_j\in A_{1}^{\ell_1}\times A_2^{\ell_2} \\ \ell_1,\ell_2=0,1, j\in [k]}} I( \bx_{\bt_{j}} \in A^{\ell_1\ell_2}) f^{A^{\ell_1\ell_2}}(\bx_{\bt_j})  
\prod_{i=1,2}p_iD(A_i^1,A_i^0). \]
Especially, when $k=2,$  we have
\[f(\bx_{\bt_1}, \bx_{\bt_2}) =\sum_{\substack{\ell_{ij}=0,1\\ i,j=1,2}}\prod_{i=1,2}I(\bx_{\bt_i}\in A^{\ell_{i1}\ell_{i2}})f(\bx_{\bt_i})\prod_{j'=1,2}\bigg(1 +\frac{(-1)^{\ell_{1j'}+\ell_{2j'}}C(|t_{1j'}-t_{2j'}|)}{\prod_{i'=1,2}p_{j'}^{\ell_{i'j'}}(1-p_{j'})^{1-\ell_{i'j'}} }\bigg), \]
 for $\bt_1,\bt_2 \in \mathbb{Z}^2$ such that  $t_{11}\neq t_{21}, t_{12}\neq t_{22}.$

The next theorem shows the characteristic function of a stationary random field $\bX=(\bX(\bt_1), \bX(\bt_2),\cdots, \bX(\bt_k))'$ constructed by (5.4-5.5).
 
 \begin{theorem} For ${ \theta}_j \in \mathbb{R}^d, \bt_j \in \mathbb{Z}^n, j=1,2,\cdots,k,$ and $ k\in \mathbb{N}/\{1\},$ 
    \begin{align*}
 E(e^{{\rm i} \sum_{j\in [k]} {\theta}_j' \bX(\bt_j)  } )= & \sum_{\substack{K_i\subset T_i, i\in [n] \\| K_i| \geq 2 \text{ or } K_i=\emptyset \\W_i\in Q_{K_i} }} 
 \sum_{\substack{A_i^0\sqcup A_i^1=T_i   \\ W_i^{\circ}\subset A_i^0 }} 
  \Big( \prod_{i\in [n]} p_i^{|A_i^1\cap K_i^c|}(1-p_i)^{|A_i^0 \cap K_i^c\cap  (W_i^\circ)^c |} (-1)^{|K_i\cap A_i^0|
}  L_i^*( W_i) \Big) \\& \times\prod_{\substack{\bt_j \in A^{D_1\times D_0}\\ D_1 \sqcup D_0 =[n] \\j\in [k] }}  E(e^{{\rm i} {\theta}_j' \bX^{A^{D_1, D_0}}  } ) , \end{align*}
where 
 $T_i=\{ t_{j,i}: j\in [k]\}, i\in [n], $  
$A^{D_1,D_0}=A^{\ell_1\cdots\ell_n }  $ where $\ell_i=1 $ for $i\in D_1$, and $\ell_i=0 $ for $i\in D_0$ for disjoint sets $D_1\sqcup D_0=[n].$
For disjoint sets $A_i^0 \sqcup A_i^1=T_i,$
 $ A^{D_1 \times D_0}$ denotes $A_1^{i_1}\times A_2^{i_2} \times \cdots \times A_n^{i_n}  $  where $i_{j}=1$ for $j\in D_1$ and $i_j=0$ for $j\in D_0.$
Also, 
 $K_i^{\circ}=\{j: j\in T_i
 /K_i, \min K_i<j<\max K_i \},$ $L_i^*(\{i_1,i_2,\cdots,i_k\})=p_i\prod_{j=2,\cdots,k}C_i^*(i_j-i_{j-1} ) $ for any $i_1<i_2<\cdots<i_k, k\in\mathbb{N}/\{1\},$
 $ C_i^*(\cdot)=C_i(\cdot)/p_i,$ and $L_i^*(\emptyset)=1.$
  \end{theorem}
  \section{Simulation}
We first provide a simulation method for GBP, then show simulation results of some stochastic models developed in previous sections to examine their empirical behaviors. Once GBP is simulated with a desired covariance, the stationary process or random field with prescribed marginal distribution is easily constructed. 

The simulation method for GBP was developed in \cite{Lee2} when the covariance function is a power function. This method can be easily modified to construct a GBP with a general covariance function developed in this paper. 
  
  GBP has ``conditioned" Markov property which means that it  loses memory before the latest observation of ``1", i.e., for any pairwise disjoint sets 
  $A,B,A',B' $  such that $\min A\cup B> \max A'\cup B',$ we have 
  \begin{align*}
&P(\cap_{i\in A} \{\xi_i=1\} \cap_{i'\in B} \{\xi_{i'}=0\} |\cap_{i\in A'} \{\xi_i=1\} \cap_{i'\in B'} \{\xi_{i'}=0\} )\\ &=P(\cap_{i\in A} \{\xi_i=1\} \cap_{i'\in B} \{\xi_{i'}=0\} | \{\xi_{i^*}=1\} \cap_{i'\in B^*} \{\xi_{i'}=0\}  )
  \end{align*}
 where $ i^*=\max A'$ and $B^*=\{i'\in B': i'> i^* \}. $
This, combined with (2.1), leads to the fact that
\begin{align*}
P(\{ \xi_{i+k}=1\} \cap_{i<i'<i+k}\{\xi_{i'}=0\} | \xi_i=1)=D(\{1, k+1\},\{2,3,\cdots, k\}),
\end{align*}
and 
\begin{align*}
P(\{ \xi_{k}=1\} \cap_{1\leq i'<k}\{\xi_{i'}=0\})=pD(\{ k\},\{1,2,\cdots, k-1\}),
\end{align*} from which the simulation method of GBP follows. 
Define for $k\in \mathbb{N},$  \[F_0(k)=\begin{dcases}
    p+\sum_{j=2}^{k} pD(\{j\}, \{1,2,\cdots,j-1\}) \} & \text{ if } k\geq 2,\\
    p& \text{ if } k=1 ,
\end{dcases}\] and
\[F(k)=\begin{dcases}
    D(\{1,2\}, \emptyset)+ \sum_{j=2}^{k} D(\{1,j+1\}, \{2,\cdots,j-1\}) \} & \text{ if } k\geq 2,\\
    D(\{1,2\}, \emptyset)& \text{ if } k=1 .
\end{dcases}\]
Now, we provide the simulation method for GBP as follows.\\ \\
 \begin{tabular}{p{.1cm} p{13cm} } 
\hline  &{\centering{ Simulation method for GBP } }\\ \hline  1.&  Generate a uniform random variable $U_1\sim \text{uniform}(0,1).$ \\ 
         2. &Let $k_1=\max\{k\in \mathbb{N}: F_0(k)\leq  U_1\}$ and  if $k_1>1,$  let $ \xi_{k_1}=1 $ and $ \xi_{j}=0$ for $1\leq j<k_1.$  If $k_1=1,$ let $\xi_1=1.$  \\ 
         3. & For $i=2,3\cdots,$ generate a uniform random variable $U_i,$ let   $k_i=\max\{k\in \mathbb{N}: F(k)\leq  U_i\}$ and  $ \xi_{i^*}=1 $ where $i^*= \sum_{j=1}^{i} k_j,$ and if $k_i\geq 2,$ let 
         $\xi_{\ell}=0$ for $\sum_{j=1}^{i-1} k_j< \ell<\sum_{j=1}^{i}k_j.$
         \\
          4. &Stop when $\sum_j^i k_j >n. $
$\{\xi_i, i=1,2,\cdots,n\}$ is a sample path of GBP.\\
         \hline 
    \end{tabular}

 After a sample path of GBP is obtained, it is easy to simulate a stationary process and random field by (5.4-5.5).

\subsection{A stationary process with a given one-dimensional marginal distribution}

We generate a stationary process, $X(1),X(2),\cdots, X(n), n=2000,$ whose one-dimensional distribution is exponential distribution, $f(x)=e^{-x},$ and covariance function is
$Cov(X(i), X(j))=c|i-j|^{2H-2}, H=.7, c=.355.$ By Example 5.8, it can be easily checked that with GBP$(p=.3, C(i)=.12|i|^{-.6})$ and  $X^A, X^{A^c}$ with $A=(-\ln(.3),\infty),$
we can construct such a process.
First, we generate GBP $\{\xi_i, i=1,\cdots, n\}$, and i.i.d. exponential random variables $Y_1,Y_2,\cdots.$ Let $A_Y=\{\ell: Y_{\ell}\in A \}, {A}^c_Y=\{\ell: Y_{\ell}\in A^c \},$   and  $X(1)=Y_{\ell_1}$ where $\ell_1= \min A_Y$ if $\xi_1=1,$ or $\ell_1= \min A^c_Y$ if $\xi_1=0.$ For $i=2,3,\cdots,n$, let $X(i)=Y_{\ell_i}$ where  $\ell_i= \min A_Y/\{\ell_1,\cdots, \ell_{i-1}\}$ if $\xi_i=1,$ or $\ell_i= \min A^c_Y/\{\ell_1,\cdots, \ell_{i-1}\} $ if $\xi_i=0.$

The graph on the top left of Figure 1 shows the simulated sample path of the stationary process with exponential marginal distribution, and the graph on the top right is the autocorrelogram of the stationary process. For comparison purposes, we provide a times series plot of i.i.d. exponential random variables and its autocorrelogram at the bottom of Figure 1. 
It is observed that our method produces a correlated sample path whose correlation decays as expected. 
\vspace{-.5cm}
\begin{figure}[h]
    \centering 
    \includegraphics[width=.8\textwidth]{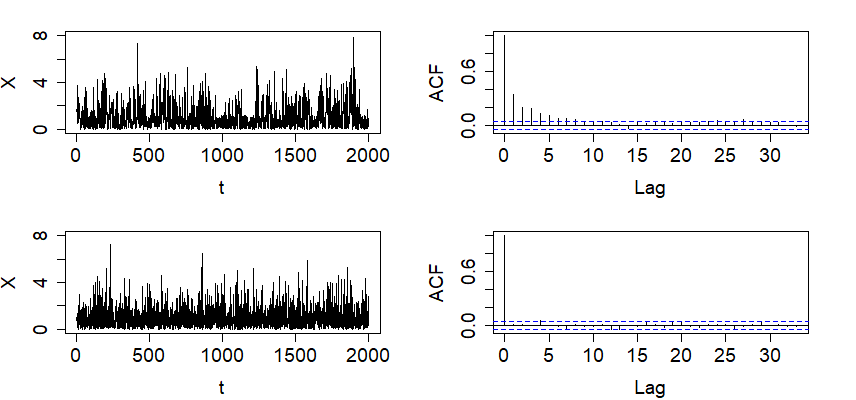}
    \caption{A correlated sample path with exponential marginal distribution and its autocorrelogram (top), a sequence of i.i.d. exponential random variables and its autocorrelogram (bottom) }
    \label{fig:enter-label}
\end{figure}\\
Since our method produces a stationary, ergodic process with an exponential marginal distribution, we expect that the empirical distribution of observations in a sample path, $\{X(i), i=1,2,\cdots, n\},$ for large $n,$  would be close to the exponential distribution. Figure 2 shows the fit of the observations in the correlated sample path to the exponential distribution.

\begin{figure}[h]
    \centering
    \includegraphics[width=.8\textwidth]{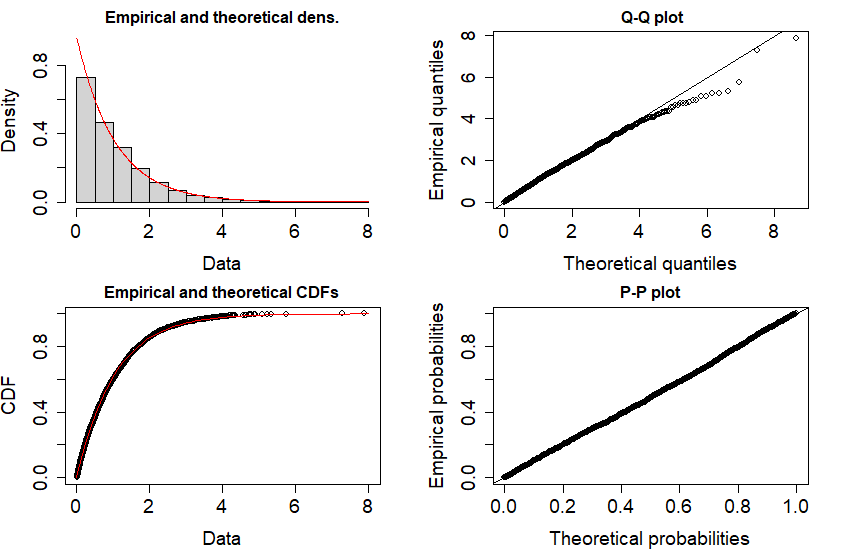}
    \caption{Comparison between the observations in the correlated sample path and exponential distribution }
    \label{fig:enter-label}
\end{figure}

%\subsection{A stationary process with Gaussian marginal distribution}
%  \begin{figure}
 %   \centering
  %  \includegraphics[width=\textwidth]{Gau.png}
   % \caption{A stationary process with Gaussian marginal distribution and power covariance function (top),  exponential covariance function (middle) \\Bottom: A sequence of i.i.d. normal random variables
    %}
    %\label{fig:enter-label}
%\end{figure}
\begin{figure}[htb]
    \centering
    \includegraphics[width=.8\textwidth]{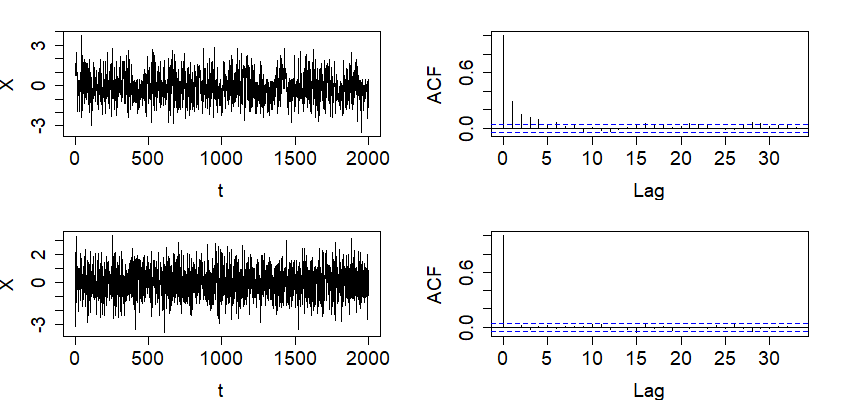}
    \caption{A correlated sample path with Gaussian marginal distribution and its autocorrelogram (top), a sequence of i.i.d. normal random variables and its autocorrelogram (bottom)}
\end{figure}
\begin{figure}[h]
    \centering
    \includegraphics[width=.8\textwidth]{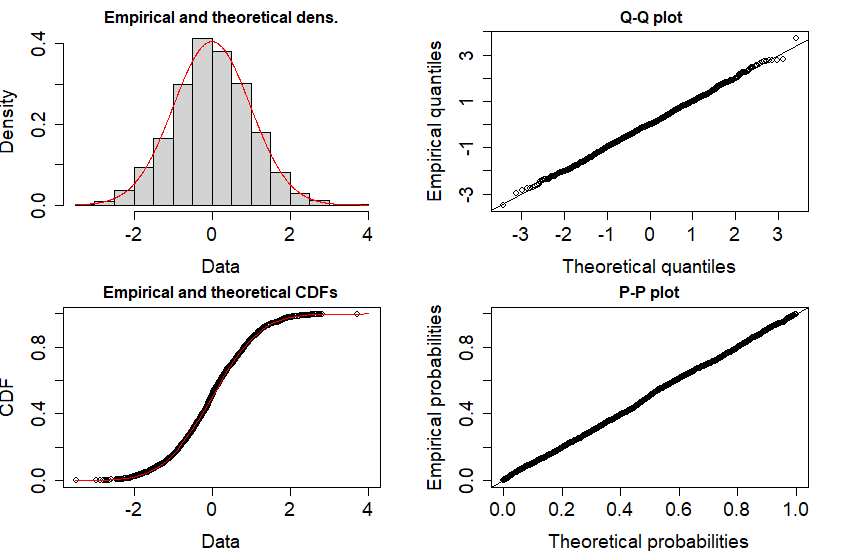}
    \caption{Comparison between the observations in the correlated sample path and standard normal  distribution}
\end{figure}
We also simulate a stationary process that has one-dimensional standard normal distribution and
the covariance function $cov(X(i), X(j))=c|i-j|^{2H-2}, H=.7, c=.329.$ It can be easily checked by Theorem 4.1 that with GBP$(p=.3, C(i)=.12|i|^{-.6} )$ and $A=(z_{.3}, \infty)$, such a process can be constructed. 
 The simulated sample path and its autocorrelogram are shown in Figure 3. For comparison purposes, we also provide the results for i.i.d. standard normal random variables on the bottom graphs in Figure 3.  The observations in the correlated sample path are compared to the standard normal distribution in Figure 4. The results indicate that our sample path shows a correlation that decays as expected and marginal Gaussian distribution.

%$The notation $z_{.3}$ is the 70th percentile of standard normal distribution. 

%(1) $f(x)=e^{-x^2}/\sqrt{2\pi}, C(i)=c|i|^{2H-2}, p=.3, H=.7, c=.12,  A=(z_{.3}, \infty), z_{.3}$=30th percentile
%\\ (2)  $f(x)=e^{-x^2}/\sqrt{2\pi}, p=.5, C(x)=.2e^{-.1x} , A=(0, \infty)$

\subsection{Multivariate stationary processes with a given marginal distribution }
We simulate a bivariate stationary process $\bX(t)=(X_1(t),X_2(t))', t=1,2,\cdots,1000,$ with a Gaussian marginal distribution, $ \bX(t)\sim N({\bf 0}, \begin{bmatrix}
    1&-.5 \\ -.5 &1 \end{bmatrix} ).$ We use GBP with $p=.258, C(i)=.2e^{-.1 |i|},$ and $A=(.2,\infty)\times(-\infty,-.2).$ 

By Theorem 4.1, the resulting process will have variance-covariance matrix
\[ corr(\bX(i),\bX(j) )= \begin{bmatrix} .386&-.386 \\ -.386 &.386 \end{bmatrix}e^{-.1 |i-j|}\]
for $i\neq j.$ 

%and  $ cov(\bX(i),\bX(i) )= \begin{bmatrix} .518&-.563 \\ -.563 &.612 \end{bmatrix}    $
%$ \int_A x_1 f(\bx)= \int_{A^c} x_2 f(\bx)=.266,   \int_A x_2 f(\bx)=\int_{A^c} x_1 f(\bx)  =-.266 $
    
The simulated sample paths are shown on the left in Figure 5, and their respective autocorrelograms are shown on the right.  The cross-correlogram is provided at the bottom in Figure 5. It is observed that the results on the correlation structures within a sample path and between sample paths are as expected.
\begin{figure}[h]
    \centering
    \includegraphics[width=.8\textwidth]{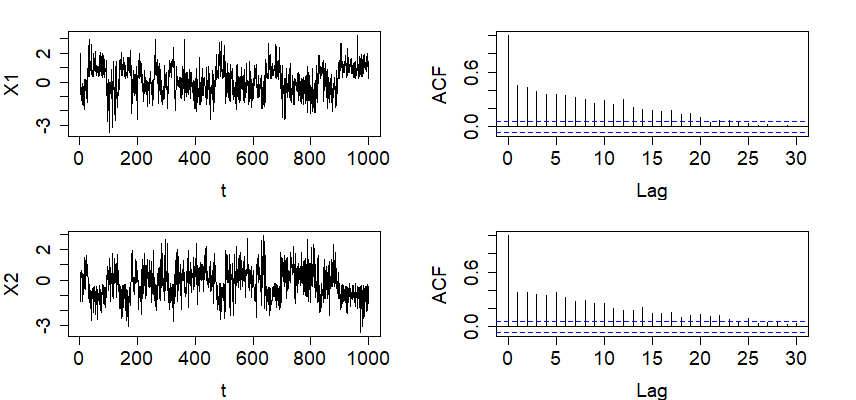}
\includegraphics[width=.35\textwidth]{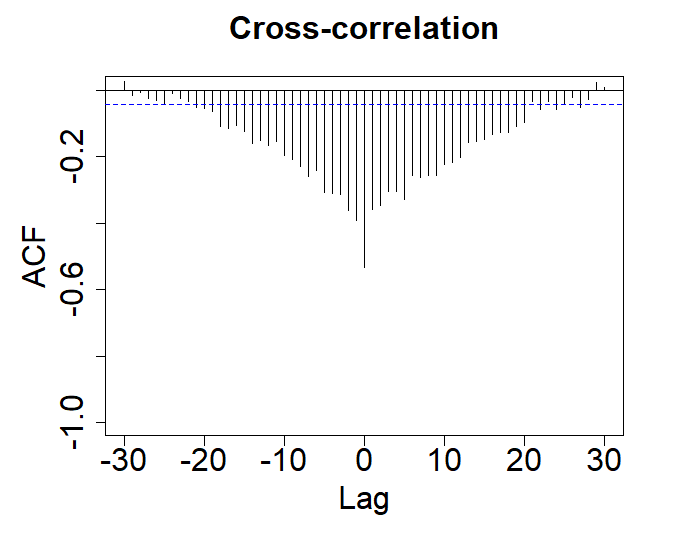}    
    \caption{A bivariate stationary process with Gaussian marginal distribution
    }
\end{figure}

\begin{figure}[h]
    \centering
    \includegraphics[width=.8\textwidth]{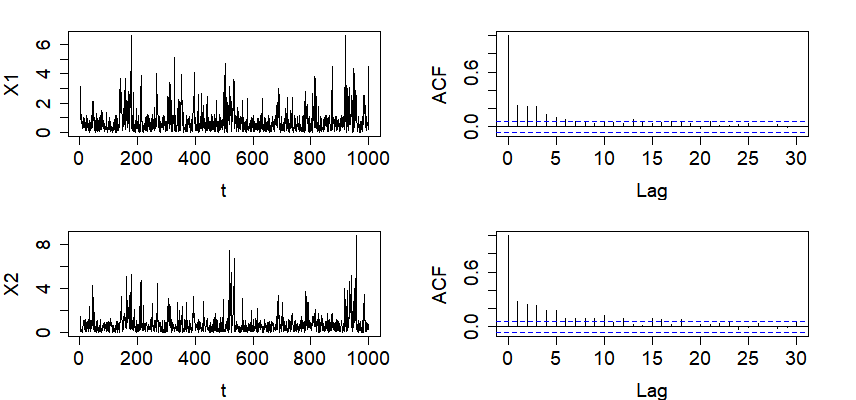}
\includegraphics[width=.35\textwidth]{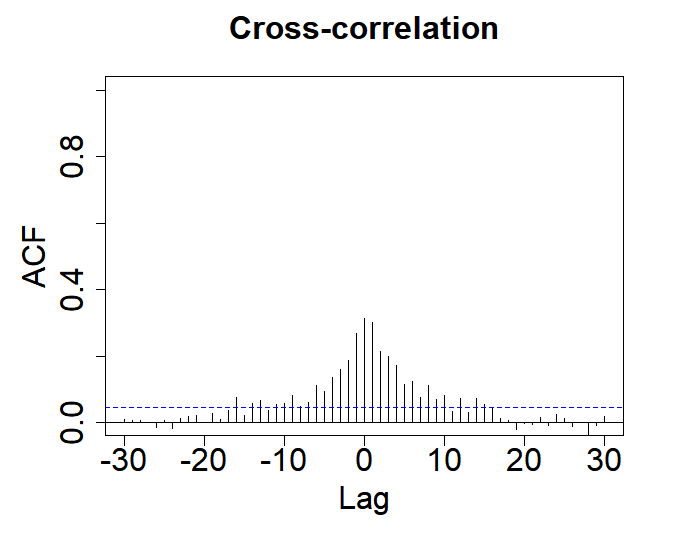}    
    \caption{A bivariate stationary process with exponential marginal distribution
    }
\end{figure}
In Figure 6, sample paths of a bivariate stationary process with exponential marginal distribution are shown. First,
a random vector of length 2 whose marginal distribution is exponential distribution is simulated with our method, i.e.,  $(X_1(t), X_2(t))'\sim f(x_1,x_2)$ where \[f(x_1,x_2)=\sum_{\ell_1,\ell_2=0,1}I(x_1\in A_0^{\ell_1}, x_2\in A_0^{\ell_2})\bigg(1 +\frac{(-1)^{\ell_1+\ell_2}c_0}{\prod_{k=1,2}p_0^{\ell_k}(1-p_0)^{1-\ell_k}}\bigg){e^{-x_1}e^{-x_2}}\] where $p_0=.3, c_0=.12, A_0^1=(-\ln{.3},\infty ),A_0^0=(0,-\ln{.3}) .$ With this marginal distribution, GBP
$(p= .339, C(i)= .12 e^{-.2 |i|})
$ and $A=\{(x_1,x_2)\in \mathbb{R}_+^2:( x_1> -\ln{.3} , 
 x_2> -\ln{.8}) \text{ or }
 ( x_1> -\ln{.8} , 
 x_2> -\ln{.3})\}
 ,$ the bivariate stationary process is simulated. Figure 6 shows its sample path, autocorrelation, and cross-correlations.
 
Since our construction method is not limited to continuous distributions, we also obtain a bivariate stationary process with binomial marginal distribution.
In Figure 7, sample paths for a bivariate stationary process $(X_1(t), X_2(t) )', t=1,2\cdots, 1000,$ with binomial marginal distribution are shown. 
The marginal distribution at $t$ is obtained from our method, 
\[f(x_1,x_2)=\sum_{\ell_1,\ell_2=0,1}I(x_1\in A_0^{\ell_1}, x_2\in A_0^{\ell_2})\bigg(1 +\frac{(-1)^{\ell_1+\ell_2}c_0}{\prod_{k=1,2}p_0^{\ell_k}(1-p_0)^{1-\ell_k} }\bigg)
f(x_1)f(x_2) \]
where $f(\cdot)$ is pmf of binomial distribution $( n=20, p=.4), p_0=.416, c_0=.12, A_0^1=\{x\in \mathbb{N}_0: x\leq 7\},A_0^0=\{x\in \mathbb{N}: 7<x\leq 20\}.$ Applying our method again with this marginal distribution, GBP $(p=.377,  C(i)=.2  e^{- .2|i|})$, and $A=\{(x_1,x_2)\in \mathbb{N}_0^2: (x_1\leq 7,x_2\leq 9) \text{ or } (x_1\leq 9,x_2\leq 7)\},$ the bivariate stationary with binomial marginal distribution is obtained. 
The sample paths, autocorrelation, and cross-correlations are shown in Figure 7.

%$p=.48, C(x)=ce^{-\theta x}, c=.2, \theta=.4$
%$A=(-z_{.4},\infty)\times(-\infty,z_{.2} ),$ and $f(\bx)=f_1(x_1)f_2(x_2)$ where $f_1,f_2$ are pdfs of standard normal distribution.

   \begin{figure}[h]
       \centering
\includegraphics[width=.8\textwidth]{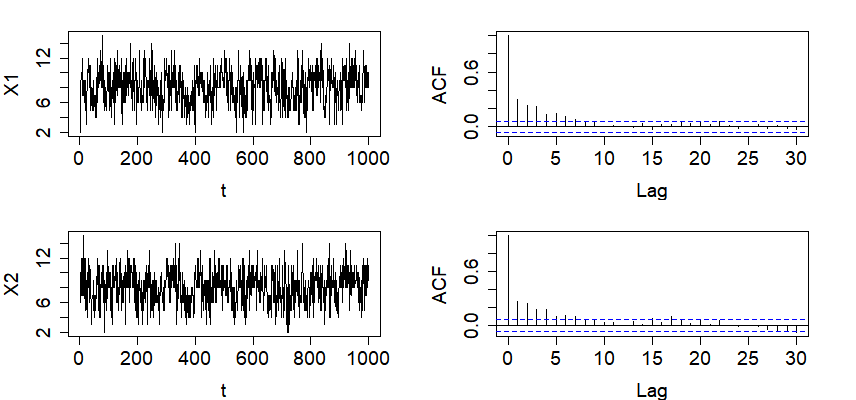}\\
\includegraphics[width=.35\textwidth]{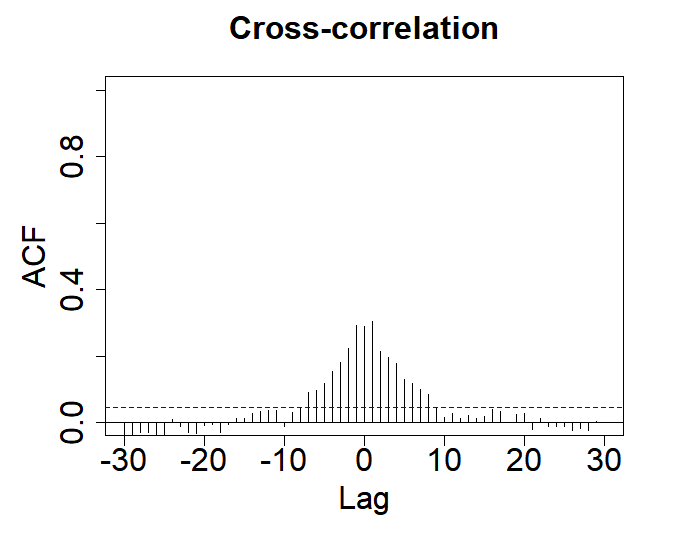}    
    \caption{A bivariate stationary process with binomial marginal distribution
    }
\end{figure}

\subsection{Stationary random fields with a given marginal distribution}
We simulate a stationary random field $X(\bt), \bt \in \{1,2,\cdots,100\}^2,$ with one-dimensional standard normal distribution.
We use GBPs with  $p_1=.4, C_1(i)=.23e^{-.4 i},$ and  $p_2=.5, C_2(i)=.24e^{-.5 i},$
and $A^{11}=(-z_{.4},z_{.4}), A^{00}=( -z_{.25},-z_{.4}) \cup (z_{.4},z_{.25}), 
A^{01}=(-\infty, -z_{.2})\cup(z_{.25},z_{.15}), A^{10}=(-z_{.2}, -z_{.25})\cup(z_{.15},\infty).$ 
It is easy to see that this satisfies (5.12-5.13), therefore, the covariance is of the form 
 in (5.14).
The simulation method follows similarly. We first simulate two independent GBPs, $\{\xi_i^j, i=1,\cdots,100\},$  with $(p_j, C_j),j=1,2, $ using the method in Table 1, and i.i.d. standard normal random variables, $Y_1,Y_2,\cdots.$ For $\bt=(1,1)',$ let  $X(\bt)=Y_{i_{\bt}} $ where $ i_{\bt}=\min A_{\bt},$  $  A_{\bt}=\{i: Y_i\in A^{\ell_1\ell_2}, \ell_j= \xi_{t_j}^j, j=1,2\}.$   For $\bt\in \{1,\cdots,100\}^2/(1,1),$ let  $X(\bt)=Y_{i_{\bt}} $ where $ i_{\bt}=\min A_{\bt}/\{i_{\bs}: (s_1\in\{1,2,\cdots, 100\} ,s_2<t_2) \text{ or } (s_1<t_2 ,s_2=t_2)\}.$

The simulated random field $X(\bt)$ is shown on the top left in Figure 8, and its estimated correlation, \[\sum_{\bt-\bt'={\bs }}\frac{(X(\bt)-\bar{X})( X(\bt')-\bar{X})}{ |N_{\bs}|}\] for $\bs\in \{-25,\cdots,-1,0,1,\cdots,25\}^2$, where $ N_{\bs}=\{ (\bt, \bt'): \bt,\bt'\in \{1,2,\cdots,100\}^2, \bt-\bt'=\bs \},$ on the top right in Figure 8.  

We also simulate a random field with one-dimensional standard normal distribution, with GBPs
    $(p_1=.5, C_1(i)=.23e^{-.4 |i|},$ $p_2=.0.549, C_2(i)=.24e^{-.5 |i|}), $ and $ A^{10}=(.5 ,1.386), A^{01}=\mathbb{R}_+/A^{10}, A^{11}=\{-x: x\in A^{01}\},
A^{00}=\{-x: x\in A^{10}\}.$  It is easy to check that  (5.9-5.10) are satisfied, therefore, the resulting random field will have the covariance of the form in (5.11).
The simulated random field and its estimated correlation are shown in the bottom graphs of Figure 8. 

\begin{figure}[htb]
    \centering
    \includegraphics[width=.35\linewidth]{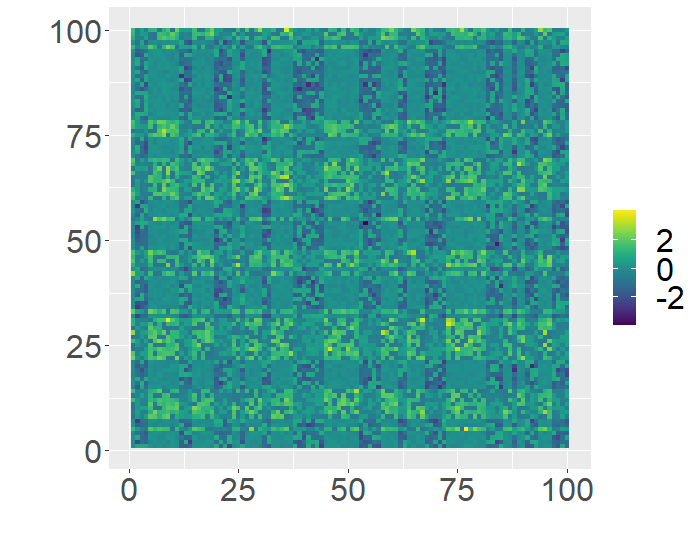}
    \includegraphics[width=.35\linewidth]{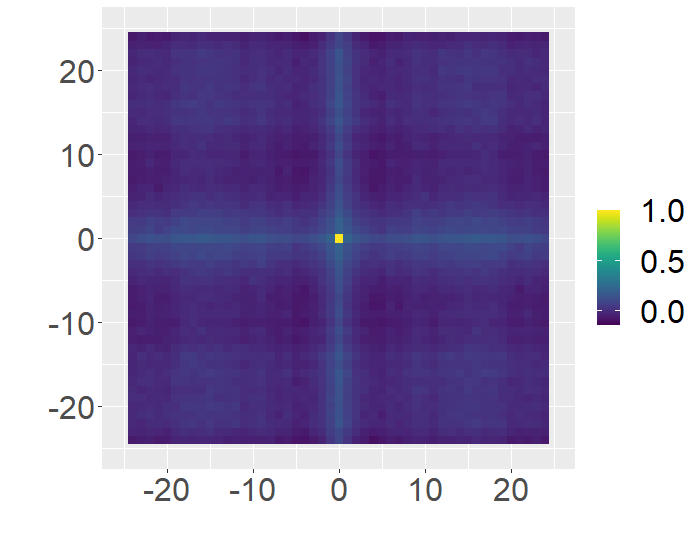}
    \includegraphics[width=.35\linewidth]{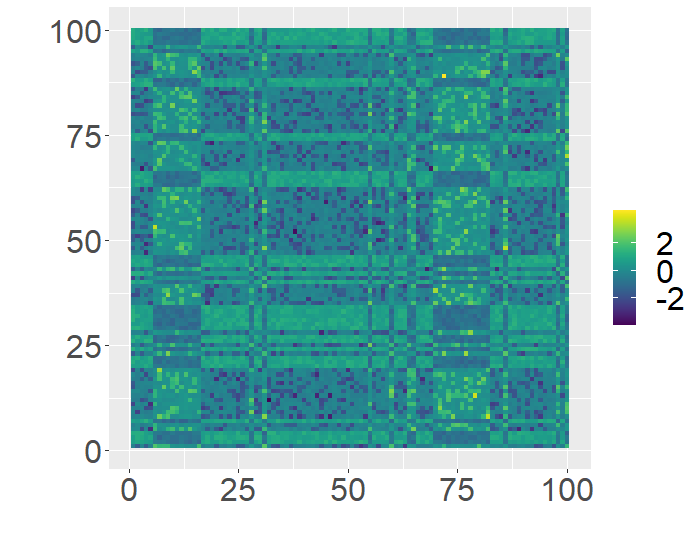}
    \includegraphics[width=.35\linewidth]{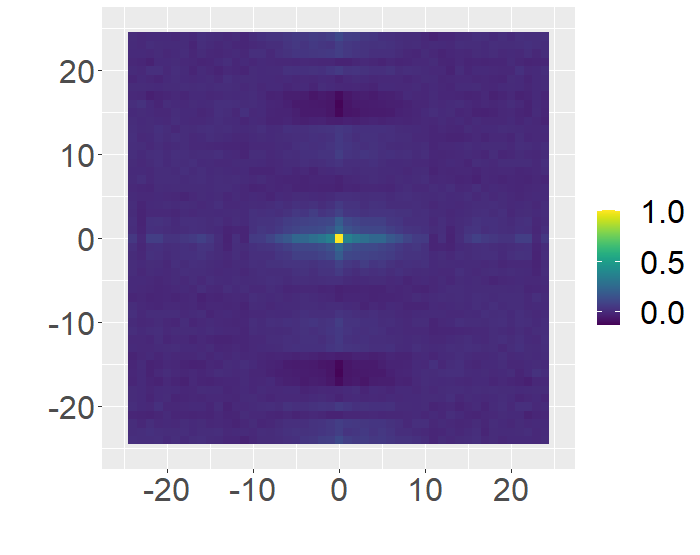}
    \caption{Random fields with one-dimensional Gaussian marginal distribution and its correlation}
\end{figure}

\section{Concluding Remarks}
We proposed a new method to construct a stationary process and random field with any one-dimensional distribution and a given covariance function that decays as fast as or slower than exponential decay.
The method utilized a generalized Bernoulli process (GBP), a stationary binary sequence with a flexible covariance structure, as a latent process. For a stationary process, a random variable was generated from one of two disjoint sets in the marginal distribution, which was chosen from the state of GBP at each time. 
The method was easily extended to a random field where several GBPs were used as latent processes, and a random variable was generated from one of several disjoint sets of the marginal distribution that is chosen from the states of the GBPs.

For future work, it would be interesting if GBP is further extended to include more broad covariance functions that can show sinuous shapes, which will enable the construction of a stochastic model with a broader covariance structure with any given marginal distribution. Also, instead of using a stationary binary sequence that has two states, one can use a stationary process with several states as a latent process to induce various dependence structures.

\section{Proofs}
\begin{lemma}
Let $C:\mathbb{N} \to \mathbb{R}_+$ be a decreasing function such that $C(x+1)/C(x)$ is non-decreasing as $x$ increase, then \[\frac{p+C(x+a)/p}{p+C(x)/p }\] increases as $x$ increases for any $a\in \mathbb{N}.$
\begin{proof}
We need to show for any $x,a\in \mathbb{N},$
\[ \frac{p+C(x+a)/p}{p+C(x)/p }< \frac{p+C(x+1+a)/p}{p+C(x+1)/p }, \]
or 
\[ \frac{p+C(x+1)/p}{p+C(x)/p }< \frac{p+C(x+1+a)/p}{p+C(x+a)/p }, \]
therefore, it is enough to show that 
\[  \frac{p+C(x+1)/p}{p+C(x)/p }< \frac{p+C(x+2)/p}{p+C(x+1)/p }\] for all $x\in \mathbb{N}.$
We will show that
\[ 2C(x+1)+C(x+1)^2/p^2<C(x)C(x+2)/p^2+C(x)+C(x+2). \]
By the assumption of $C,$ it follows that $ C(x+1)^2\leq C(x)C(x+2).$
Also, \begin{align*}
    C(x+1)^2 <(C(x)/2-C(x+2)/2)^2 +C(x)C(x+2) =(C(x)/2+C(x+2)/2)^2,
\end{align*}
therefore, $  2C(x+1)< C(x)+C(x+2), $
and the result follows.
\end{proof}
\end{lemma}

\begin{lemma} Under Assumption 2.3, 
$D(A,\{i\} )=L(A)-L(A\cup \{i\}) >0$ for any $A, \{i\} \subset \mathbb{N}, i\notin A.$

    \begin{proof}
    If $A=\emptyset,$ $D(A,\{i\} )=1/p-1>0.$
    If $A=\{i'\} , D(A,\{i\} )=1-(p+C^*( |i'-i| ) )  >0 $ by Assumption 2.3 $\it ii).$
    
    Let $|A|=2.$ We will show that
    $D(\{i_0,i_2\} ,\{i_1\} )=L(\{i_0,i_2\})-L(\{i_0,i_1, i_2\}) >0$ for any $i_0,i_1,i_2 \in \mathbb{N}.$ 
    
    If $ \max\{i_0,i_2\}<i_1$ or $ \min\{i_0,i_2\}>i_1,$ the result easily follows since
    \[ L(\{i_0,i_1, i_2\})= \begin{dcases}
        L(\{i_0,i_2\})(p+C^*(|i_1-\min\{i_0,i_2\}  |)) & \text{if } i_1 <\min\{i_0,i_2\} ,\\
        L(\{i_0,i_2\}) (p+C^*(|i_1-\max\{i_0,i_2\}  |)) & \text{if } i_1 >\max\{i_0,i_2\}.
    \end{dcases} \]

Now, assume $i_0<i_1<i_2.$ 
We need to show that \[ \frac{L(\{i_0,i_2 \})}{L(\{i_0,i_1,i_2\})}=\frac{p+C^*(x+a) }{(p+C^*(x) )(p+C^*(a))  } >1,\] for any $x=i_1-i_0 ,a=i_2-i_1 \in \mathbb{N}.$ 
By Lemma 8.1, 
\[ \frac{p+C^*(x+a) }{(p+C^*(x) )(p+C^*(a))  } \geq  \frac{p+C^*(1+a) }{(p+C^*(1) )(p+C^*(a))  } \geq \frac{p+C^*(2) }{(p+C^*(1) )(p+C^*(1))  }>1  .\] 

For $|A|>2, $ it is derived in the same way as $|A|=2.$ 
    If $ \max A<i$ or $ \min A>i,$ 
    \[ L(A\cup \{i\})= \begin{dcases}
        L(A)(p+C^*(|i-\min A  |)) & \text{if } i <\min A ,\\
        L(A) (p+C^*(|i-\max A  |)) & \text{if } i >\max A,
    \end{dcases} \] therefore, $D(A, \{i\}) >0.$
If $\min A <i < \max A,$ let $i^*=min\{j: j\in A, j>i \} $, $i_*=max\{j: j\in A, j<i \}. $
Then, \[\frac{L(A)}{L(A\cup \{i\}) }= \frac{L(\{ i_*, i^* \})}{L(\{i_*, i, i^* \}) }>1. \] 

    \end{proof}
\end{lemma}

%\begin{lemma}[Lemma 5.1 in \cite{Lee}]
%For any $\{i_0,i_1,\cdots,i_n\} \subset \mathbb{N},$
%\begin{align*}
%&D(\{i_0,i_{n}\},\{ i_1,i_2,\cdots,i_{n-1}\}) =\\&  D(\{i_0,i_{n}\},\{ i_1,i_2,\cdots,i_{n-2}\})- D(\{i_0,i_{n-1},i_{n}\},\{ i_1,i_2,\cdots,i_{n-2}\})   .\end{align*}
%In general, for any $\{a_0,a_1,\cdots, a_n, b_0,b_1,\cdots, b_m \} \subset \mathbb{N},$
%\begin{align}
%&D(\{a_0,a_1,\cdots, a_n\},\{ b_0,b_1,\cdots, b_m \})=D(\{a_0,a_1,\cdots, a_n\},\{ b_0 \})\nonumber \\&-D(\{a_0,a_1,\cdots, a_n,b_1\},\{ b_0\})- D(\{a_0,a_1,\cdots, a_n,b_2\},\{ b_0,b_1\}) \nonumber\\&\cdots  -D(\{a_0,a_1,\cdots, a_n, b_{m}\},\{  b_0,b_1,\cdots, b_{m-1}\}).
%\end{align} 
%\end{lemma}

%\begin{lemma}[Lemma 5.2 in \cite{Lee}] For any $\{a_0, a_1, \cdots, %a_n, a_0',a_1',\cdots, a_n'\} \subset \mathbb{R}_{+}$ such that
%$a_0-\sum_{i=1}^{j}a_i>0, a_0'-\sum_{i=1}^{j}a_i'>0,$ for %$j=1,2,\cdots,n$, if
%\[\frac{a_0}{a_0'} >\frac{a_1}{a_1'}>\cdots>\frac{a_n}{a_n'}, \] then \[ \frac{a_0-a_1-a_2-\cdots -a_n}{a_0'-a_1'-a_2'-\cdots -a_n'}> \frac{a_0}{a_0'}.\]
%\end{lemma}

{\bf {Proof of Theorem 3.1.}}\\
We will prove for any disjoint sets $A_1\sqcup A_0\subset \mathbb{Z}$
\begin{equation}
    D(A_1, A_0)>0
\end{equation}by mathematical induction. 
If $A_0=\emptyset, D(A_1,A_0)=L(A_1)>0.$ Assume $A_0\neq \emptyset.$

If $|A_0|=1,$ $ D(A_1,A_0)=L(A_1) - L(A_1\cup\ A_0 )>0,$ by Lemma 8.2. 
Assume (8.1) holds with $|A_0|\leq m.$
We will show that (8.1) holds when $|A_0|=m+1.$ Let $A_0=\{i_0', i_1', i_2' , \cdots, i_m'\}$ with $i_0'<i_1'<i_2'<\cdots<i_m'.$ By definition of $D$, it follows that
\begin{align*}
D(A_1,A_0)&=  \sum_{j=0}^{|A_0|-1}\sum_{ \substack{F'\subset A_0\setminus \{i_m'\}\\ |F'|=j}}(-1)^{j} L(A_1\cup F')+\sum_{j=0}^{|A_0|-1}\sum_{ \substack{F''\subset A_0\setminus \{i_m'\}\\ |F''|=j}}(-1)^{j+1} L(A_1\cup F''\cup\{i_m'\})\\&=  D(A_1,A_0\setminus \{ i_m'\}) -
  D(A_1\cup \{i_m'\}, A_0\setminus\{ i_m'\}),\end{align*}  
therefore,
it is enough to show that \[  D(A_1,A_0\setminus \{ i_m'\}) >
  D(A_1\cup \{i_m'\}, A_0\setminus\{ i_m'\})  .\]
By
Lemma 5.1 in \cite{Lee}, 
\begin{equation}
    \frac{D(A_1,A_0\setminus\{ i_m'\})}{  D(A_1\cup\{i_m'\},A_0\setminus\{ i_m'\}) } =\frac{a_0-a_1-a_2-a_3\cdots-a_{m-1}}{ a_0'-a_1'-a_2'-a_3'\cdots-a_{m-1}'} , \end{equation}
where 
\begin{align*}
{a_{j}}= \begin{dcases} {  D(A_1\cup \{i_{j}'\},\{ i_0',i_1',\cdots,i_{j-1}'\}) } &\text{for } j=1,\cdots, m-1,\\ {  D(A_1,\{i_0'\}) } &\text{for } j=0, \end{dcases} \end{align*} and
\begin{align*}
{a_{j}'}= \begin{dcases} {   D(A_1\cup\{i_{j}',i_{m}'\},\{ i_0',i_1',\cdots,i_{j-1}'\}) } &\text{for } j=1,\cdots, m-1,\\ {   D(A_1\cup\{i_{m}'\},\{ i_0'\}) }, &\text{for } j=0. \end{dcases}
\end{align*}
We will apply Lemma 5.2 {\it i)} in \cite{Lee}  to show that $(8.2)>1.$ Note that $ a_0-a_1-\cdots -a_{j} >0$ 
and 
$ a_0'-a_1'-\cdots -a_{j}' >0$ for $j=0,1,\cdots,m-1,$ by the earlier assumption that (8.1) holds for $|A_0|\leq m.$

Define $i_j^*= \max \{ i_j', i: i\in A_1, i<i_m' \}.$ Note that $i_j^*$ is non-decreasing as $j$ goes from 1 to $m-1$. If  $\max A_1 >i_m' $, define $i^*= \min \{i: i\in A_1, i>i_m' \}.$ 
 For $j=1,\cdots, m-1,$
\[\frac{a_{j}}{a_{j}'}=\frac{  D(A_1\cup \{i_{j}'\},\{ i_0',i_1',\cdots,i_{j-1}'\}) }{  D(A_1\cup\{i_{j}',i_{m}'\},\{ i_0',i_1',\cdots,i_{j-1}'\}) }= \frac{\sum_{\ell=0}^{j}\sum_{ \substack{F'\subset  \{i_0',i_1',\cdots,i_{j-1}'\}\\ |F'|=\ell}}(-1)^{\ell} L(A_1\cup\{i_j'\}\cup F')}{\sum_{\ell'=0}^{j}\sum_{ \substack{F''\subset \{ i_0',i_1',\cdots,i_{j-1}' \}\\ |F''|=\ell'}}(-1)^{\ell'} L(A_1\cup\{i_j', i_m'\} \cup F'')} .\]
Note that
if $A_1=\emptyset,$ then, 
for any $F'=F''\subset  \{i_0',i_1',\cdots,i_{j-1}'\},$
\[ \frac{L(A_1\cup\{i_j'\} \cup F')}{  L(A_1\cup\{i_j', i_m'\} \cup F'')}  = \frac{1}{p+C^*(i_{m}'-i_j') }.\]
Similarly, for any $F'=F''\subset  \{i_0',i_1',\cdots,i_{j-1}'\},$ if  $\max A_1 <i_m'$, then
\[ \frac{L(A_1\cup\{i_j'\} \cup F')}{  L(A_1\cup\{i_j', i_m'\} \cup F'')}  = \frac{1}{p+C^*(i_{m}'-i_j^*) },\]
and if $\max  A_1 >i_m',$
then
\[ \frac{L(A_1\cup\{i_j'\} \cup F')}{  L(A_1\cup\{i_j', i_m'\} \cup F'')}  = \frac{p+C^*(i^*-i_j^*)}{(p+C^*(i_{m}'-i_j^*))(p+C^*(i^*-i_m' )) } .\]
Therefore,
\begin{align}
\frac{a_{j}}{a_{j}'} =\begin{dcases}\frac{p+C^*(i^*-i_j^*)}{(p+C^*(i_{m}'-i_j^*))(p+C^*(i^*-i_m' )) } &\text{ if }  \max  A_1 >i_m',  \\ \frac{1}{p+C^*(i_{m}'-i_j^*) } & \text{ if }   \max A_1 <i_m',\\
\frac{1}{p+C^*(i_{m}'-i_j') } & \text{ if }   A_1=\emptyset,
\end{dcases}  \end{align}  which is non-increasing as $j$ goes from 1 to $m-1,$ since
$p+C^*(x)$ is a decreasing function and $(p+C^*(x+a))/(p+C^*(x)) $   is  a non-decreasing function of $x$ for any $a>0$ by Lemma 8.1. Also,
\begin{equation*}
  \frac{a_{0}}{a_{0}'}=\frac{  D(A_1,\{i_0'\}) }{   D(A_1\cup\{i_{m}'\},\{ i_0'\}) }=\frac{L(A_1)-L(A_1\cup \{i_0'\})}{ L(A_1\cup \{i_m'\})- L(A_1\cup \{i_0', i_m'\})  }
\end{equation*}
with
\begin{equation*}
    \frac{L(A_1)}{L(A_1\cup \{i_m'\})}=\begin{dcases}
    \frac{1}{p+C^*(i^*-i_{m}') } &\text{ if } \min A_1>i_m',
    \\ \frac{1}{p+C^*(i_{m}'-i_*) } &\text{ if } \max A_1<i_m' ,\\
  \frac{p+C^*(i^*-i_*)}{(p+C^*(i^*-i_{m}'))(p+C^*(i_m'-i_*)) } &\text{ if } \min A_1< i_m'<\max A_1, \\
   1/p&\text{ if } A_1=\emptyset,
    \end{dcases}
\end{equation*} where $i_*=\max\{i: i\in A_1, i<i_m'\}$ if $\min A_1<i_m',$ and 

\begin{equation*}
 \frac{L(A_1\cup \{i_0'\})}{ L(A_1\cup \{i_0', i_m'\})  }=(8.3)
\end{equation*} with $j=0.$
Since $i_*\leq i_0^*\leq i_1^* \leq i_2 ^* \leq \cdots \leq i_{m-1}^*,$
\begin{align*}
  &\frac{L(A_1)}{ L(A_1\cup \{i_m'\})}  \geq \frac{L(A_1\cup \{i_0'\})}{ L(A_1\cup \{i_0', i_m'\})  } \geq  \frac{a_1}{a_1'}  \geq  \frac{a_2}{a_2'} \geq \cdots   \geq  \frac{a_{m-1}}{a_{m-1}'}
\end{align*}
therefore, by Lemma 5.2  {\it i)} in \cite{Lee},
\[(8.3)\geq  \frac{L(A_1)}{ L(A_1\cup \{i_m'\})}. \]
The result follows as $\frac{L(A_1)}{ L(A_1\cup \{i_m'\})}>1$ by Lemma 8.2. 
\qed \\ \\

{\bf Proof of Theorem 4.1.}\\
  Note that for any set $B \subset \mathbb{R}^d$ and any $i\in \mathbb{Z},$
  \begin{align*}
      P(\bX(i) \in B)&= P(\bX^A(i)\in B )P(\xi_i=1)+
  P(\bX^{A^c}(i)\in B )P(\xi_i=0)\\
&=\frac{\int_{A\cap B} f({\bf x}) d{\bf x}}{p}p +\frac{\int_{A^c \cap B} f({\bf x}) d{\bf x}}{1-p}(1-p)= \int_B f({\bf x})d{\bf x}.
  \end{align*}
  Also, \[P( \cap_{j=1}^k \{\bX({i_j})\in B_j\})=P(  \cap_{j=1}^k \{\bX({i_j+\ell})\in B_j    ) \] 
  for any $ k \in \mathbb{N} , \ell, i_1,i_2,\cdots,i_k \in \mathbb{Z}, $
since $\{\xi_i, i\in \mathbb{Z}\}$ is stationary, $\{\bX^A(i), i\in \mathbb{Z}\}$ and $ \{\bX^{A^c}(i), i\in \mathbb{Z}\}$ are i.i.d., and the three sequences of random variables are independent of each other.
  Therefore, $\{\bX(i), i\in \mathbb{Z}\}$ is a stationary process with marginal pdf $f.$
      Without loss of generality, let $\mu=0.$ If $\mu\neq0,$ one can use $\bX(i)-\mu= (\bX^A(i) -\mu) I_{\{\xi_i =1\}} +(\bX^{A^c}(i)-\mu) I_{\{\xi_i =0\}}. $ Since
      \begin{align*}
          E(X_k(i)X_{\ell}(j))&=\sum_{u, v=0,1}E(X_k^{A^u})E(X_{\ell}^{A^v}) E(I_{\{\xi_i =u\}}  I_{\{\xi_j =v\}} )
          \\&= \sum_{u, v=0,1} \frac{\int_{A^u} x_kf({\bf x}) d{\bf x}\int_{A^v} x_{\ell}f({\bf x}) d{\bf x} \Big( p^{u+v}(1-p)^{2-u-v} +(-1)^{u+v}C(|i-j|) \Big)}{p^{u+v}(1-p)^{2-u-v}}
       % \\&=d_kd_{\ell}\Big(1+\frac{C(|i-j|)}{p^2}\Big)-2d_kd_{\ell}\Big(1-\frac{C(|i-j|)}{p(1-p)} \Big) +d_kd_{\ell}\Big(1+ \frac{C(|i-j|)}{(1-p)^2} \Big)
      \\&=\frac{d_kd_{\ell} }{p^2(1-p)^2} C(|i-j|),
      \end{align*}
      where $A^0=A^c,A^1=A$, and $d_k=\int_A x_kf({\bf x}) d{\bf x}$, the result follows.
  \qed \\ \\
{\bf Proof of Proposition 4.3.}\\
      We can show that
      \begin{align*}
          &P(\{\bX(i)\in B_1\} \cap \{\bX(j)\in B_2\})=  
       \Big( \prod_{k=1,2}  P(\bX^{A}(i)\in B_k )  \Big)  P(\{\xi_i =1\} \cap \{\xi_j =1\} )\\&+\Big(P(\bX^A(i)\in B_1 )P(\bX^{A^c}(i)\in B_2 )+P(\bX^A(i)\in B_2 )P(\bX^{A^c}(i)\in B_1 )\Big) P(\{\xi_i =0\} \cap \{\xi_j =1\} )\\&+\Big( \prod_{k=1,2}P(\bX^{A^c}(i)\in B_k ) \Big) P(\{\xi_i =0\} \cap \{\xi_j =0\} )
          \\ &=\Bigg( \prod_{k=1,2} \frac{\int_{A\cap B_k} f({\bf x}) d{\bf x}}{p}\Bigg) p(p+C^*(|i-j|))\\& + \bigg( \frac{\int_{A\cap B_1} f({\bf x}) d{\bf x}\int_{A^c\cap B_2} f({\bf x}) d{\bf x}}{p(1-p)}+ \frac{\int_{A\cap B_2} f({\bf x}) d{\bf x}\int_{A^c\cap B_1} f({\bf x}) d{\bf x}}{p(1-p)}\Bigg) 
 p(1-p-C^*(|i-j|))\\&+ \Bigg( \prod_{k=1,2} \frac{\int_{A^c\cap B_k} f({\bf x}) d{\bf x}}{1-p}\Bigg) ( (1-p)^2+pC^*(|i-j|))
\\&=\prod_{k=1,2} \Bigg( \int_{A\cap B_k} f({\bf x})d{\bf x} + \int_{A^c \cap B_k} f({\bf x})d{\bf x} \Bigg) + 
 pC^*(|i-j|)  \prod_{k=1,2} \Bigg( \frac{\int_{A\cap B_k} f({\bf x}) d{\bf x}}{p} - \frac{\int_{A^c \cap B_k } f({\bf x}) d{\bf x}}{1-p} \Bigg) . \end{align*}
     Since $P(\bX(i) \in B_k)= \int_{A\cap B_k} f({\bf x})d{\bf x} + \int_{A^c \cap B_k} f({\bf x})d{\bf x} ,$ for $ k=1,2,$ the result follows.
  \qed \\ \\

  {\bf Proof of Theorem 4.6.}\\
Here, we abuse the notations and understand $L^*(K), D(B,F)$ for $K, B, F \subset \mathbb{Z}$
as $ L^*(\{i_j: j\in K\})$ and
$ D(\{i_j: j\in B\}, \{i_j:j\in F \})$, respectively.
 By (3.1), 
\begin{align*}
    &E(e^{i \sum_{j=1,2,\cdots,k} \theta_j \bX(i_j)  } )= \sum_{\substack{B\cup F= \{1,2,\cdots,k\} \\ B\cap F=\emptyset }} \prod_{i\in B}E(e^{{\rm i} \theta_i'\bX^A }) \prod_{j\in F}E(e^{{\rm i} \theta_j'\bX^{A^c} })pD(B,F)
    \\&= (G)+(H)\end{align*} where 
    (G)=terms that does not include  $L^*$, and  (H)=terms that include $L^*(K)$.

In $pD(B,F),$ the terms that do not include $L^*$ are by the definition of the operations $D, L,$ \[
    \sum_{j=0}^{|F|}\sum_{ \substack{F'\subset F\\ |F'|=j}}(-p)^{j} p^{|B|}=p^{|B|}(1-p)^{|F|} ,\] 
therefore,
\begin{align*}
  (G)=&\sum_{\substack{B\cup F= \{1,2,\cdots, k\} \\ B\cap F=\emptyset }} \prod_{i\in B}E(e^{{\rm i} \bt_i'\bX^A }) \prod_{j\in F}E(e^{{\rm i} \bt_j'\bX^{A^c} }) p^{|B|}(1-p)^{|F|} =\prod_{j=1}^k \Big(E(e^{{\rm i} \bt_j'\bX^A }) p + E(e^{{\rm i} \bt_j'\bX^{A^c} })(1-p)  \Big) \\&=\prod_{j=1}^kE(e^{{\rm i} \bt_j'\bX }).\end{align*} 
 The terms that  include $L^*(K)$ in $pD(B,F)$ are
 \[\sum_{\substack{K\subset [k]\\ |K|\geq 2 } }\sum_{ F'\subset F} \sum_{\substack{ 
  K\subset (B\cup F')\\ W\in Q_K \\  W^{\circ} \cap (B\cup F')=\emptyset  }} (-1)^{|F'|} p^{|B|+|F'|-|K|} L^*(W)
    \]
    where $W^{\circ}=\sqcup K_i^{\circ}$ for $W=\{K_i,i=1,2,\cdots,|W|\},$
therefore,

\begin{align*}
 (H)= &\sum_{\substack{K\subset [k]\\ |K|\geq 2 \\W\in Q_K}}\prod_{j\in W^{\circ}}E(e^{{\rm i} \theta_j'\bX^{A^c} }) \sum_{\substack{B\cup F= \{1,\cdots, k\}/W^{\circ} \\ B\cap F=\emptyset  }} \prod_{i\in B}E(e^{{\rm i} \theta_i'\bX^A }) \prod_{\ell\in F}E(e^{{\rm i} \bt_{\ell}'\bX^{A^c} }) \\&\sum_{\substack{F'\subset F \\ K\cap B^c \subset  F'}}
  p^{|B|}(-p)^{|K\cap B^c |}(-p)^{|F'|-|K\cap B^c |}p^{-|K|}L^*(W)
  \\&=\sum_{\substack{K\subset [k]\\ |K|\geq 2 \\W\in Q_K}}\prod_{j\in W^{\circ}}E(e^{{\rm i} \theta_j'\bX^{A^c} }) \sum_{\substack{B\cup F= \{1,\cdots, k\}/W^{\circ} \\ B\cap F=\emptyset }} \prod_{i\in B}E(e^{{\rm i} \theta_i'\bX^A }) \prod_{\ell\in F}E(e^{{\rm i} \theta_{\ell}'\bX^{A^c} }) \\&
  p^{|B|}(-p)^{|K\cap B^c |}(1-p)^{|F|-|K\cap B^c |}p^{-|K|}  L^*(W)
\\&  = \sum_{\substack{K\subset [k]\\ |K|\geq 2 \\W\in Q_K}}\prod_{j\in W^{\circ}}E(e^{{\rm i} \theta_j'\bX^{A^c} }) \sum_{\substack{B\cup F= \{1,\cdots, k\}/W^{\circ} \\ B\cap F=\emptyset }} \prod_{i\in B}E(e^{{\rm i} \theta_i'\bX^A }) \prod_{\ell\in F}E(e^{{\rm i} \theta_{\ell}'\bX^{A^c} }) \\&
  p^{|B\cap K^c|}(-1)^{|K\cap B^c |}(1-p)^{|F|-|K\cap B^c |} L^*(W) \\& = \sum_{\substack{K\subset [k]\\ |K|\geq 2 \\W\in Q_K}}\prod_{j\in W^{\circ}}E(e^{{\rm i} \bt_j'\bX^{A^c} }) \prod_{i\in K}(E(e^{{\rm i} \theta_i'\bX^A })  -  E(e^{{\rm i} \theta_i'\bX^{A^c} }) ) \\& \prod_{\ell \in \{1,\cdots,k\}/(W^{\circ}\cup K) } (pE(e^{{\rm i} \theta_{\ell}'\bX^A })+ (1-p)E(e^{{\rm i} \theta_{\ell}'\bX^{A^c} }) ) L^*(W).\end{align*}
\qed \\ \\

{\bf Proof of Theorem 5.1.}\\
  It is easy to see that $\{\bX(\bt), \bt\in \mathbb{Z}^2 \}$ is a stationary random field with marginal pdf $f.$
  Also,
\begin{align}
E(\bX(\bt) \bX(\bs)'  )&=\sum_{\ell_1,\ell_2,\ell_1', \ell_2'=0,1}
\bd_{\ell_1 \ell_2} \bd_{\ell_1' \ell_2'}' P( \{\xi_{t_1}^1=\ell_1\}\cap \{\xi_{s_1}^1=\ell_1'\})P({\{\xi_{t_2}^2=\ell_2\} \cap\{\xi_{s_2}^2=\ell_2'\}})  
 \nonumber \\&=\sum_{i,i',j,j'=0,1}\bd_{ij} \bd_{i'j'}' \Big(p_1^{i+i'}(1-p_1)^{2-i-i'}+(-1)^{i+i'}C_1(|t_1-s_1|) \Big) \nonumber \\& \times \Big(p_2^{j+j'}(1-p_2)^{2-j-j'}+(-1)^{j+j'}C_2(|t_2-s_2|)\Big) 
 \nonumber \\&=\Big(\sum_{\ell_1,\ell_2=0,1} \bd_{\ell_1\ell_2}\prod_{k=1,2} p_k^{ \ell_k} (1-p_k)^{ 1-\ell_k}\Big)\Big(\sum_{\ell_1,\ell_2=0,1} \bd_{\ell_1\ell_2}\prod_{k=1,2} p_k^{ \ell_k} (1-p_k)^{ 1-\ell_k} \Big)'\nonumber\\&+ p_1p_2C_1^*(|t_1-s_1|)C_2^*(|t_2-s_2|){\bf v}_0{\bf v}_0' +p_1C_1^*(|t_1-s_1|) {\bf v_1}{\bf v_1}'
\nonumber\\&+p_2C_2^*(|t_2-s_2|){\bf v_2}{\bf v_2}' \nonumber
\end{align}  
where
\begin{align*}
 \bd_{\ell_1 \ell_2}&= E(\bX^{A^{\ell_1\ell_2}}(\bt))=\Bigg(\frac{ \int_{A^{\ell_1\ell_2}}  x_1 f( \bx ) d\bx  }{\prod_{k=1,2} p_k^{ \ell_k} (1-p_k)^{ 1-\ell_k} }, \cdots, \frac{ \int_{A^{\ell_1\ell_2}}  x_d f( \bx ) d\bx  }{\prod_{k=1,2} p_k^{ \ell_k} (1-p_k)^{ 1-\ell_k} } \Bigg)',  \end{align*} for $\ell_1,\ell_2=0,1, $ and 
${\bf v_0}= \bd_{10}+\bd_{01} -\bd_{11}-\bd_{00}     ,$
$  {\bf v_1}=p_2(\bd_{11}-\bd_{01}) +(1-p_2)(\bd_{10}-\bd_{00}), $
$ {\bf v_2}=p_1(\bd_{11}-\bd_{10}) +(1-p_1)(\bd_{01}-\bd_{00}). $
Since \[\sum_{\ell_1,\ell_2=0,1} \bd_{\ell_1\ell_2}\prod_{k=1,2} p_k^{ \ell_k} (1-p_k)^{ 1-\ell_k}=E(\bX(\bt)),\] the result follows. For other cases, the results are derived similarly.
\qed \\ \\ 

 {\bf Proof of Theorem 5.4.}\\ 
  First, assume that $t_j\neq s_j$ for all $j=1,2,\cdots, n,$ i.e., $O_{(\bt-\bs)}=
  \emptyset.$ Then,
  \begin{align}
E(\bX({\bf t}) \bX({\bf s})')=\sum_{\substack{\ell_1,\ell_2, \cdots, \ell_n=0,1\\ \ell_1',\ell_2', \cdots, \ell_n'=0,1 } } E\big(\bX^{A^{\ell_1\cdots \ell_n}}\big)   E\big(\bX^{A^{\ell_1'\cdots \ell_n'}}\big)'\prod_{j\in [n]} P(\xi_{t_j}^j=\ell_j ,\xi_{s_j}^j=\ell_j' ) .
  \end{align}
Note that $P(\xi_{t_j}^j=\ell_j ,\xi_{s_j}^j=\ell_j' )= p_j^{\ell_j}(1-p_j)^{1-\ell_j} p_j^{\ell_j'}(1-p_j)^{1-\ell_j'}  + (-1)^{2-(\ell_j+\ell_j' )}C_j(|t_j-s_j|)$ by (3.1), which leads to
  \begin{align*}
    (8.4)&=\Bigg(\sum_{\ell_1,\cdots, \ell_n=0,1 } E\big(\bX^{A^{\ell_1\cdots \ell_n}}\big)   \prod_{j\in [n] }p_j^{\ell_j}(1-p_j)^{1-\ell_j} \Bigg) \Bigg(
      \sum_{\ell_1',\cdots, \ell_n'=0,1 }
    E\big(\bX^{A^{\ell_1'\cdots \ell_n'}}\big)'  \prod_{j\in [n] } p_j^{\ell_j'}(1-p_j)^{1-\ell_j'}\Bigg)\\&+
      \sum_{\substack{K\subset [n]\\ K \neq \emptyset }} \prod_{k\in K}
      C_k(|t_k-s_k|)
     \Bigg( \sum_{\ell_1,\cdots, \ell_n=0,1 } (-1)^{|K|-\sum_{i\in K} \ell_i}E\big(\bX^{A^{\ell_1\cdots \ell_n}}\big)   \prod_{j\in [n]/K }p_j^{\ell_j}(1-p_j)^{1-\ell_j} \Bigg)\\&
     \Bigg( \sum_{\ell_1',\cdots, \ell_n'=0,1 }
     (-1)^{|K|-\sum_{i\in K} \ell_i'}  E\big(\bX^{A^{\ell_1'\cdots \ell_n'}}\big)'  \prod_{j\in [n]/K } p_j^{\ell_j'}(1-p_j)^{1-\ell_j'}\Bigg).
  \end{align*}
Since
  \[  E(\bX)= \sum_{\ell_1,\cdots, \ell_n=0,1 } E\big(\bX^{A^{\ell_1\cdots \ell_n}}\big)   \prod_{j\in [n]}p_j^{\ell_j}(1-p_j)^{1-\ell_j} ,  \]
 the result follows.

 If $O_{(\bt-\bs)} \neq \emptyset,$
 (8.4) becomes
 \[\sum_{\substack{\ell_{j'}=0,1\\j'\in O_{(\bt-\bs)}}} \Bigg(\prod_{j'\in O_{(\bt-\bs)}}P(\xi_{t_{j'} }^{j'}=\ell_{j'} ) \Bigg) \sum_{\substack{\ell_j,\ell_j'=0,1 \\ j\in [n]/O_{(\bt-\bs)} } } E\big(\bX^{A^{\ell_1\cdots \ell_n}}\big)   E\big(\bX^{A^{\ell_1'\cdots \ell_n'}}\big)'\prod_{i\in [n]/O_{(\bt-\bs)} } P(\xi_{t_i}^i=\ell_i ,\xi_{s_i}^i=\ell_i' ) ,\]
 and similarly to the previous case, we can show that
 \begin{align*}
&\sum_{\substack{\ell_j,\ell_j'=0,1 \\ j\in [n]/O_{(\bt-\bs)} } } E\big(\bX^{A^{\ell_1\cdots \ell_n}}\big)   E\big(\bX^{A^{\ell_1'\cdots \ell_n'}}\big)'\prod_{i\in [n]/O_{(\bt-\bs)} } P(\xi_{t_i}^i=\ell_i ,\xi_{s_i}^i=\ell_i' )
\\&={\bf q q'}+  \sum_{\substack{K\subset [n]/O_{(\bt-\bs)} \\ K \neq \emptyset }} \bigg(\prod_{k\in K}
      C_k(|t_k-s_k|)\bigg) {\bf v v'} \end{align*} where
 \begin{align*}
     {\bf q}&=\sum_{\substack{\ell_j=0,1 \\ j\in [n]/O_{(\bt-\bs)} } }E\big(\bX^{A^{\ell_1\cdots \ell_n}}\big)   \prod_{i\in [n]/O_{(\bt-\bs)} }p_i^{\ell_i}(1-p_i)^{1-\ell_i},
\\ {\bf v}&=
 \sum_{\substack{\ell_{i'}=0,1 \\ i'\in [n]/O_{(\bt-\bs)} } } (-1)^{|K|-\sum_{i\in K} \ell_i}E\big(\bX^{A^{\ell_1\cdots \ell_n}}\big)   \prod_{j\in [n] /(K\cup O_{(\bt-\bs)}) }p_j^{\ell_j}(1-p_j)^{1-\ell_j} ,
 \end{align*}
 from which the result follows.
 \qed \\ \\

 {\bf Proof of Proposition 5.5.}\\ 
 For $k\subset K', K\neq \emptyset,$ we have from (5.7)
 \begin{align*}
     {\bf m}_K=\sum_{\substack{\ell_j=0,1\\  j\in K/ O_{(\bt-\bs)}}}(-1)^{O^{K/ O_{(\bt-\bs)} }} \frac{\int_{A(\{\ell_i, i\in K \cup O_{(\bt-\bs)} \}) } x_{i'} f(\bx) d\bx} {\prod_{i\in K \cup O_{(\bt-\bs)} }p_i^{\ell_i}(1-p_i)^{1-\ell_i}}= (1-1)^{|K/O_{(\bt-\bs)}|}\mu_{i'}=0,
 \end{align*}
 where $O^{K/O_{(\bt-\bs)}}$ denotes the 
 number of zeros in $\ell_j, j\in K/O_{(\bt-\bs)}$.
\qed \\ \\

{\bf Proof of Lemma 5.7.}\\ 
We assume that $A= supp(f)=\mathbb{R}.$ General cases can be proved in a similar way. 
Let $p=1/2^k$ for some $k\in \mathbb{N}.$ 
One can find disjoint intervals $r_i \subset \mathbb{R}, i=1,\cdots, 2^k $ such that $\mathbb{R}=\sqcup_{i=1,\cdots, 2^k} r_i $ and $\int_{r_i} f(x)dx=p.$ If $\int_{r_i} g(x)f(x)dx = p \int g(x)f(x)dx$ for some $i$, then we found $A_1=r_i.$ 

If $\int_{r_j} g(x)f(x)dx < p \int g(x)f(x)dx$ for some $j,$ then there is $j'$ such that $\int_{r_{j'}} g(x)f(x)dx > p \int g(x)f(x)dx,$ since otherwise   $\int g(x)f(x)dx=\sum_{i=1}^{2^k}\int_{r_i} g(x)f(x)dx < 2^k p \int g(x)f(x)dx=  \int g(x)f(x)dx.$ Let $\sup r_i=S_i$ and $ \inf r_i=I_i$ for $i=1,\cdots, 2^k.$ Assume $S_j<S_{j'},$ for $S_j>S_{j'}$ it is proved in a similar way.  For $y \in (S_j,S_{j'}), $ define an interval $r^*_y=(y_b ,y)$ such that  $ y_b \in  (I_j,I_{j'}) $ and $\int_{r^*_y} f(x) dx=p .$  Let $G({y})=\int_{r^*_y} g(x) f(x) dx .$ Then, $G(\cdot)$ is a continuous function with $\lim_{y\downarrow S_j }G(y)= \int_{r_j} g(x)f(x)dx $ and $ \lim_{y\uparrow S_{j'} }G(y)= \int_{r_{j'}} g(x)f(x)dx ,$ therefore, there should be $z\in (S_j,S_{j'})$ such that $ G(z)=p\int g(x) f(x) dx.$ Then we found $A_1=r_z^*.$ 

If $A\subsetneq supp(f),$ $r_i$ can be replaced by $s_i=r_i\cap supp(f)$ where $r_i, i=1,2\cdots,$ are disjoint intervals such that $\int_{s_i} f(x) dx=p,  $ and the result follows in the same way. If $supp(f)=\mathbb{R}^d,$ we consider disjoint sets $\sqcup_{i=1,\cdots, 2^k} \mathbb{R}^{d-1}\times r_i=\mathbb{R}^d$
where $r_i, i=1,\cdots, 2^k,$ are disjoint intervals, and the proof follows in the same way.

For any $p\in (0,1),$ the result also holds as follows. Note that $
p=\sum_{k\in B}1/2^k$ for some $B\subset \mathbb{N},$ and for each $k\in B,$ we can find $r_{z_k}^*$ such that $ \int_{r_{z_k}^*} f(x) dx=1/2^k $ and $\int_{r_{z_k}^*}g(x) f(x) dx=1/2^k \int g(x) f(x) dx.$ Let $A_1=\cup_{k\in B}  r_{z_k}^*,$ then we have $\int_{A_1} f(x) =p$ and $\int_{A_1} g(x)f(x) dx=p \int g(x)f(x) dx   $ by the dominated convergence theorem.
\qed \\ \\

{\bf Proof of Theorem 5.12.}\\
 Let $\bt_j=(t_{j,1},t_{j,2},\cdots, t_{j,n})', $
 and 
 $T_i=\{ t_{j,i}: j=1,2,\cdots,k\}, $ for $i=1,2,\cdots,n, j=1,2,\cdots, k.$
 Also, for disjoint sets $D_1\sqcup D_0=\{1,2,\cdots,n\},$ define
 $A^{D_1,D_0}=A^{\ell_1\cdots\ell_n }  $ where $\ell_i=1 $ for $i\in D_1$, and $\ell_i=0 $ for $i\in D_0.$
 Also, for disjoint sets $A_i^0 \sqcup A_i^1=T_i,  i=1,2\cdots, n,$
let $ A^{D_1 \times D_0}=A_1^{i_1}\times A_2^{i_2} \times \cdots \times A_n^{i_n}  $  where $i_{j}=1$ for $j\in D_1$ and $i_j=0$ for $j\in D_0.$

\begin{align*}
 &E(e^{{\rm i} \sum_{j\in [n]} {\theta}_j' \bX(\bt_j)  } )= \sum_{\substack{A_i^0\sqcup A_i^1=T_i   \\ i\in [n] }}
 \prod_{\substack{\bt_j \in A^{D_1\times D_0}\\ D_1 \sqcup D_0 =[n]\\j\in [k
 ]  }} E(e^{{\rm i} {\theta}_j' \bX^{A^{D_1, D_0}}  } ) \prod_{i\in [n]} p_iD_i( A_i^1, A_i^0)
 \\&=(G)+(H) \end{align*}
 where $(G)=$ terms that do not include $L^*,$ and $(H)=$ terms that include $L^*.$
 It is easy to see that 
 \begin{align*}
(G)&= \sum_{\substack{A_i^0\sqcup A_i^1=T_i,   \\ i\in [n] }}
 \prod_{\substack{\bt_j \in A^{D_1\times D_0}\\ D_1 \sqcup D_0 =[n]\\j\in [k
 ]  }} E(e^{{\rm i} {\theta}_j' \bX^{A^{D_1, D_0}}  } ) \prod_{i\in [n]} p_i^{|A_i^1|}(1-p)^{|A_i^0|} 
 \end{align*}
 
Also,
 \begin{align*}
 (H)&= \sum_{\substack{K_i\subset T_i, i\in [n] \\| K_i| \geq 2 \text{ or } K_i=\emptyset \\ \cup K_i\neq \emptyset}} \sum_{W_i\in Q_{K_i} }
 \sum_{\substack{A_i^0\sqcup A_i^1=T_i,  \\ W_i^{\circ}\subset A_i^0 }} 
\prod_{\substack{\bt_j \in A^{D_1\times D_0}\\ D_1 \sqcup D_0 =[n] \\j\in [k] }}  E(e^{{\rm i} {\theta}_j' \bX^{A^{D_1, D_0}}  } ) \times \\& \Big( \sum_{\substack{A_j^{00}\subset A_j^0\cap K_j^c\cap  {(W_j^{\circ})}^c \\ j\in [n]}}\prod_{i\in [n]} p_i^{|A_i^1\cap K_i^c|}(-p_i)^{|A_i^{00}|} (-1)^{|K_i\cap A_i^0|
}\Big) 
 \prod_{i\in [n]} L_i^*( W_i) \\&= \sum_{\substack{K_i\subset T_i, i\in [n] \\| K_i| \geq 2 \text{ or } K_i=\emptyset\\ \cup K_i\neq \emptyset  }} \sum_{W_i\in Q_{K_i}}
 \sum_{\substack{A_i^0\sqcup A_i^1=T_i,   \\ W_i^{\circ}\subset A_i^0 }} 
  \Big( \prod_{i\in [n]} p_i^{|A_i^1\cap K_i^c|}(1-p_i)^{|A_i^0 \cap K_i^c\cap  {(W_i^{\circ})}^c |} (-1)^{|K_i\cap A_i^0|} \Big)  \prod_{i\in [n]} L_i^* (W_i) 
  \\& \prod_{\substack{\bt_j \in A^{D_1\times D_0}\\ D_1 \sqcup D_0 =[n] \\j\in [k] }} E(\exp( {\rm i} {\theta}_j' \bX^{{D_1, D_0}} ) ),\end{align*}
therefore, the result follows.
\qed \\ \\

%%%%%%%%%%%%%%%%%%%%%%%%%%%%%%%%%%%%%%%%%%%%%%
%% Single Appendix:                         %%
%%%%%%%%%%%%%%%%%%%%%%%%%%%%%%%%%%%%%%%%%%%%%%
%\begin{appendix}
%\section*{???}%% if no title is needed, leave empty \section*{}.
%\end{appendix}
%%%%%%%%%%%%%%%%%%%%%%%%%%%%%%%%%%%%%%%%%%%%%%
%% Multiple Appendixes:                     %%
%%%%%%%%%%%%%%%%%%%%%%%%%%%%%%%%%%%%%%%%%%%%%%
%\begin{appendix}
%\section{???}
%
%\section{???}
%
%\end{appendix}

%%%%%%%%%%%%%%%%%%%%%%%%%%%%%%%%%%%%%%%%%%%%%%
%% Support information, if any,             %%
%% should be provided in the                %%
%% Acknowledgements section.                %%
%%%%%%%%%%%%%%%%%%%%%%%%%%%%%%%%%%%%%% %%%%%%%%

%\hspace{-.5cm}{\bf Acknowledgments}:
%The author would like to thank the referees for their valuable comments that improved the manuscript.\\
%{\bf Funding information:}
%The authors states no funding involved.
%\\
%{\bf Author contribution:}
%The author confirms the sole responsibility for the conception of the study, presented results and
%manuscript preparation.
%\\ 
%{\bf Conflict of interest:}
%The author states no conflict of interest.

%%%%%%%%%%%%%%%%%%%%%%%%%%%%%%%%%%%%%%%%%%%%%%
%% Funding information, if any,             %%
%% should be provided in the                %%
%% funding section.                

%%%%%%%%%%%%%%%%%%%%%%%%%%%%%%%%%%%%%%%%%%%%%%
%% Supplementary Material, including data   %%
%% sets and code, should be provided in     %%
%% {supplement} environment with title      %%
%% and short description. It cannot be      %%
%% available exclusively as external link.  %%

\end{document}